\documentclass[aps,eqsecnum,preprint,prd,showpacs]{revtex4}
\usepackage{graphicx}
\def\bold#1{\setbox0=\hbox{$#1$}%
     \kern-.025em\copy0\kern-\wd0
     \kern.05em\%\baselineskip=18ptemptcopy0\kern-\wd0
     \kern-.025em\raise.0433em\box0 }
\def\slash#1{\setbox0=\hbox{$#1$}#1\hskip-\wd0\dimen0=5pt\advance
         to\wd0{\hss\sl/\/\hss}}

\newcommand{\be}{\begin{equation}}
\newcommand{\ee}{\end{equation}}
\newcommand{\bea}{\begin{eqnarray}}
\newcommand{\eea}{\end{eqnarray}}
\newcommand{\nn}{\nonumber}

\begin{document}
\begin{titlepage}
\addtolength{\jot}{10pt}

 \preprint{\vbox{\hbox{BARI-TH/06-534 \hfill}
                \hbox{April 2006\hfill} }}
 
\vspace*{1cm}                

\title{\bf Exclusive $B \to K^{(*)} \ell^+ \ell^-$,
$B \to K^{(*)} \nu {\bar \nu}$ and $ B \to K^* \gamma$ transitions \\in a
scenario with a single Universal Extra Dimension \\}

\author{P. Colangelo$^a$, F. De Fazio$^a$, R. Ferrandes$^{a,b}$, T.N. Pham$^c$ \\}

\affiliation{ $^a$ Istituto Nazionale di Fisica Nucleare, Sezione di Bari, Italy\\ 
$^b$ Dipartimento di Fisica,  Universit\'a  di Bari, Italy\\ 
$^c$ Centre de Physique Th\'eorique, \\
Centre National de la Recherche Scientifique, UMR 7644\\
\'Ecole Polytechnique, 91128 Palaiseau Cedex, France \\}

\begin{abstract}
We analyze the exclusive  rare  $B \to K^{(*)} \ell^+ \ell^-$, $B
\to K^{(*)} \nu \bar \nu$ and $B \to K^* \gamma$ decays in the
Applequist-Cheng-Dobrescu model, which is an extension of the
Standard Model in presence of universal extra dimensions. In particular, we
consider the case of a single universal extra dimension. We study how spectra, branching fractions and
asymmetries depend on  the compactification parameter $1/R$, and whether the hadronic
uncertainty due to the  form factors obscures or not such a dependence. We find that, using present data,
it is possible in many cases to put a sensible lower bound to  $1/R$,  the most
stringent one coming from $B \to K^* \gamma$. We also study how improved experimental data
can be used to establish stronger constraints to this model.
 \end{abstract}


\maketitle
\end{titlepage}

\newpage
\section{Introduction} \label{sec:intro}
Although the Standard Model (SM) of electroweak interactions has
successfully passed several experimental tests, it is commonly
believed that a more fundamental theory should exist. Direct
evidence of new Physics will be hopefully gained at high energy
colliders such as the LHC. In the meanwhile, signals of
new interactions and particles can be obtained indirectly through
the analysis of processes which are rare or even forbidden in the
Standard Model. Among these, rare B decays induced by $b \to s$
transition play a peculiar role since they are  induced at loop
level and hence they are suppressed in the Standard Model.

Present data show that such a suppression indeed occurs.  In the
case of $b \to s \gamma$ modes, which in SM are induced by the
electromagnetic penguin diagrams dominated by top and W  exchange, the
branching fractions  have been measured both for inclusive and
exclusive transitions; we collect them in Table \ref{brbsgamma}.

\begin{table}[h]
    \caption{Branching fractions  of rare B decays induced by
    $b \to s \gamma$ transition.}
    \label{brbsgamma}
    \begin{center}
    \begin{tabular}{cclcr}
    \hline  \hline
Mode & Belle Collab. & &BaBar Collab. &\\ \hline
$B \to X_s\gamma$ & $(3.55 \pm 0.32 \pm^{0.30}_{0.31} \pm^{0.11}_{0.07})
\times 10^{-4}$&\cite{Koppenburg:2004fz}&  $(3.27 \pm 0.18
\pm^{0.55}_{0.40}\pm^{0.04}_{0.09})\times 10^{-4}$&
\cite{Aubert:2005cu}  \\ \hline
 $B^0 \to K^{*0} \gamma$ &  $(4.01 \pm
0.21 \pm 0.17 ) \times 10^{-5}$& \cite{Nakao:2004th} &
 $(3.92 \pm
0.20\pm 0.24 )
 \times 10^{-5}$& \cite{Aubert:2004te}
\\ \hline
$B^- \to K^{*-} \gamma$ & $(4.25 \pm 0.31 \pm 0.24)
\times 10^{-5}$ & \cite{Nakao:2004th}&  $(3.87 \pm 0.28 \pm
0.26 ) \times 10^{-5}$&  \cite{Aubert:2004te}
\\ \hline \hline
    \end{tabular}
  \end{center}
\end{table}

Modes  with two leptons in the final
state, such as $B \to X_s \ell^+ \ell^-$,  $B \to K^{(*)} \ell^+
\ell^-$ and $B \to X_s \nu \bar \nu$, $B \to K^{(*)} \nu \bar \nu$,  are also suppressed. 
$b \to s \ell^+ \ell^-$ transitions are described in SM by QCD, magnetic
and electroweak semileptonic penguin operators, which give rise to
an effective Hamiltonian composed of ten operators, as we shall
see in more detail below. The resulting SM predictions depend on
both the chirality structure of such operators, both on the value
of their Wilson coefficients. The situation is simpler in the case
of $b \to s \nu {\bar \nu}$ modes, described by $Z$ penguin and
box diagram dominated by top exchange: the corresponding effective
Hamiltonian is composed by a single operator, therefore
new Physics effects can  just  induce an operator with opposite
chirality or modify the value of the  Wilson
coefficient, a  scenario  relatively simple to analyse.

From the experimental point of view, the most recent measurements of the branching fractions are
provided by Belle \cite{Iwasaki:2005sy,Abe:2004ir} and BaBar 
\cite{Aubert:2004it,Aubert:2005cf} Collaborations and are collected in Table
\ref{brbsleptons}.
\begin{table}[t]
    \caption{Branching fractions  of  rare B decays induced by
    $b \to s \ell^+ \ell^-$ and $b \to s \nu {\bar \nu}$ transitions.}
    \label{brbsleptons}
    \begin{center}
    \begin{tabular}{ c c l c r }
    \hline  \hline
Mode & Belle Collab. & & BaBar Collab. &  \\ \hline
$B \to X_s \ell^+ \ell^- $ & $ (4.11  \pm 0.83
\pm^{0.85}_{0.81}) \times 10^{-6}$&\cite{Iwasaki:2005sy}\,\,\,\,\,\,\,\,&
$(5.6 \pm 1.5 \pm 0.6 \pm 1.1) \times 10^{-6}$ \,\,& \cite{Aubert:2004it}   \\
\hline $B \to K^* \ell^+ \ell^- $ & $(16.5 \pm^{2.3}_{2.2} \pm 0.9
\pm  0.4)\times 10^{-7}$ &\cite{Abe:2004ir} & $(7.8
\pm^{1.9}_{1.7} \pm 1.2)\times
      10^{-7}$& \cite{Aubert:2005cf} \\ \hline
      $B\to K \ell^+ \ell^- $ & $(5.50 \pm^{0.75}_{0.70} \pm 0.27 \pm
0.02) \times 10^{-7}$ &\cite{Abe:2004ir} & $(3.4 \pm 0.7 \pm
0.3) \times 10^{-7}$ &\cite{Aubert:2005cf} \\
\hline $B \to X_s \nu \bar \nu $ & & & & \\ \hline
$B \to K^* \nu \bar \nu $ &  & & & \\ \hline
$B^- \to K^- \nu \bar \nu $ & $<3.6 \times 10^{-5} \,\,(90 \% \,\, CL)$  &\cite{Abe:2005bq} &
 $<5.2 \times 10^{-5} \,\,\,\,(90 \% \,\, CL) $ &\cite{Aubert:2004ws} \\ \hline \hline
\end{tabular}
\end{center}
\end{table}
%
In addition, important information can be gained from the
 forward-backward lepton
asymmetry  in $B \to K^* \ell^+ \ell^-$, which
is a powerful tool to distinguish between  SM
and several extensions of it. Belle Collaboration has recently
provided the first measurement of such an asymmetry
\cite{Abe:2005km}.

Among the various models of Physics beyond the SM, those with
extra dimensions are attracting  interest for
manifold reasons. For example, they provide a unified framework
for gravity and other interactions, giving some hints on the
hierarchy problem and a connection with string theory.
Particularly interesting are the scenarios with {\it universal}
extra dimensions, in which  all the SM fields are allowed to
propagate in the extra dimensions. Their feature is that  compactification
of the extra dimensions leads to the
appearance of Kaluza-Klein (KK) partners of the SM fields in the
four dimensional description of the higher dimensional theory, together with
 KK modes without  corresponding SM partners. A simple scenario is represented by  the
Appelquist-Cheng-Dobrescu (ACD) model \cite{Appelquist:2000nn}
with a single compactified extra dimension, which presents the
appealing feature of introducing  only one additional free
parameter with respect to the Standard Model, i.e. $1/R$, the
inverse of the compactification radius. 

Analyses aimed  at identifying the signatures 
of extra dimensions in processes already
accessible at particle accelerators or within the reach of future facilities 
give different bounds to the sizes of extra dimensions, depending on the 
specific model considered. The bounds are more severe in the case of UED, and in the
5-d scenario   constraints from Tevatron run I allow to put  the bound
$1/R \ge 300 $ GeV. In the following  we  analyze a broader
range  $1/R \geq 200$ GeV to be more general.

Rare $B$ transitions can be used to constrain the ACD scenario 
\cite{Agashe:2001xt}.
In particular, Buras and collaborators 
have investigated  the impact of universal extra
dimensions  on the $B^0_{d,s}-\bar B^0_{d,s}$ mixing mass differences,  on
the CKM unitarity triangle  and on inclusive 
$b \to s$ decays for which  they have computed the 
 effective Hamiltonian  \cite{Buras:2002ej,Buras:2003mk}. 
 The availability of  precise data on  exclusive $b \to s$-induced decays, 
collected in Tables    \ref{brbsgamma},\ref{brbsleptons},  induces us to extend  the analysis 
to  such  processes in the framework of   the
 ACD model. In this case, the uncertainty in the hadronic form factors must be taken
 into account, since it can  overshadow the sensitivity to the compactification 
 parameter. Actually,  we show that this is not the case, at least  for the smallest values of $1/R$:
 computing, for example, the  branching
 fractions of $B \to K^{(*)} \ell^+  \ell^-$ as well as the forward-backward lepton asymmetry 
 in $B \to K^* \ell^+  \ell^-$ for a representative set of form factors we find that a bound can 
 be put, and it can be improved following the improvements in the experimental data.
Moreover, since in the limit of large $1/R$ the Standard Model is recovered, we  also
investigate  the  agreement of current data with  SM predictions
\cite{Hurth:2003vb}.
We have  also considered the decays $B \to K^{(*)} \nu \bar \nu$, 
although for these modes no signal has been observed,
so far, studying how observables like
the various helicity amplitudes for  $B \to K^*$ transitions
are  modified in the ACD model. Finally, we have considered
 the branching fraction of $B \to K^* \gamma$ as a
 function of  $1/R$,  pointing out that it allows  to 
 establish the most stringent  bound for  the compactification parameter.

 The plan of the paper is as follows: in the next Section we
 briefly describe the ACD model.  We discuss the  modes $B\to K^{(* )}\ell^+ \ell^-$, $B\to K^{(* )}\nu \bar \nu$  and $B\to K^* \gamma$   in the
 subsequent Sections; finally, we present our conclusions and the perspectives for further analyses.

\section{Models with extra dimensions: the ACD model with a single UED} \label{sec:acd}
If  other dimensions exist in our universe
apart the usual 3 spatial + 1 temporal ones, and if such extra dimensions are compactified,
fields living in all dimensions would manifest themselves in the 3+1 space by the appearence of
Kaluza-Klein excitations (from the original  Kaluza and
Klein studies aimed at unifying electromagnetism and gravity by the introduction of one
 extra dimension  \cite{KK}).  For example, in the case of a single extra dimension with coordinate
 $x_5=y$ compactified on a
circle of radius R (the compactification radius), a  field $F(x,y)$ (with $x$ denoting the whole of the
usual 3+1 coordinates) would be a
 periodic function of $y$,  and hence it could be expressed as
 \be F(x,y)=\sum_{n=- \infty}^{n=+  \infty} F_n(x) e^{i\, n \cdot y/R} \,\,\, .\ee 
If, for example, $F$ is  a boson field obeying the  equation of motion 
 $\partial_M \partial^M F(x,y)=0$ (M=0,1,2,3,5),
the KK modes $F_n$ would obey the equation
 \be \left(\partial^\mu
\partial_\mu+{n^2 \over R^2} \right)F_n(x)=0 \,\,\, \hskip 1cm
 \mu=0,1,2,3\,\,\,, \ee  and therefore, apart the zero mode, they would
 behave in four dimensions as massive  particles with $m_n^2= (n/R)^2$. 

An important question is whether ordinary  fields  
propagate  or not in all the dimensions. One possibility  is that
only  gravity  propagates in  the whole ordinary + extra
dimensional Universe, the "bulk". 
Opposed to this are  models with {\it
universal} extra dimensions (UED), in which all the fields 
 propagate in all available dimensions.

In this paper  we  focus on the model developed by Appelquist,
Cheng and Dobrescu (ACD) \cite{Appelquist:2000nn} that  belongs to
the UED scenarios. It  consists in the minimal extension of the
SM in $4+ \delta$ dimensions, and we  consider the simplest
case $\delta=1$. This extra dimension is compactified to the orbifold $S^1/Z_2$,
with the coordinate $x_5=y$ running from 0 to $ 2 \pi R$. The points $y=0, y=\pi R$
are fixed points of the orbifold; the boundary conditions at these points determine
the Kaluza Klein mode expansion of the fields. 
Under the parity transformation $P_5: y \to -y$ fields having a correspondent in 
 the 4-d SM should be even, so that their zero mode in the KK mode expansion can be interpreted
as the ordinary SM field. On the other hand, fields having
no correspondent in the SM should be odd,  and therefore they do not have zero modes in the KK
expansion. For example,  in this scenario a vector
field has  a fifth component which is
odd under $P_5$.

Important features of the ACD model are: i) there is
 a single  additional free parameter with respect to the SM, 
the compactification radius $R$; ii) the conservation
of KK parity, which has the consequence that there is no tree-level
contribution of KK modes in low energy processes (at scales 
 $\mu \ll 1/R$) and no production of single KK excitation in ordinary particle
 interactions.
 
 A detailed description of the
 extension of  SM in five dimensions is provided in
\cite{Buras:2002ej}; here we recall the main features of such a
construction.

\begin{itemize}

\item{\bf Gauge group}

The gauge bosons associated to   $SU(2)_L \times U(1)_Y$ gauge group
are $W_M^a$ ($a=1,2,3$, $M=0,1,2,3,5$) and  $B_M$, and the
 gauge couplings are: ${\hat g}_2=g_2 \sqrt{2 \pi R}$
and ${\hat g}^\prime=g^\prime \sqrt{2 \pi R}$
(we denote with the caret the quantities
referring to the five dimensional description). The charged bosons are 
$W_M^\pm={1 \over \sqrt{2}}(W_M^1 \mp W_M^2)$.  Moreover,
 as in  SM, $W_M^3$ and $B_M$ mix
giving
rise to the fields $Z_M$ and $A_M$. The mixing angle is
defined through the ordinary relations: \be c_W=\cos \theta_W={{\hat g}_2
\over \sqrt{{\hat g}_2^2+{\hat g}^{\prime 2}}}\hskip 1 cm s_W=\sin
\theta_W={{\hat g}^\prime \over \sqrt{{\hat g}_2^2+{\hat
g}^{\prime 2}}} \,\,\, . \label{weinberg} \ee Due to the relations between
the five and four dimensional gauge constants, 
the Weinberg angle is the same as in  SM. 
On the other hand, the gauge bosons associated to $SU(3)_c$ are the gluons $G^a_M(x,y)$
 ($a=1,..,8$).

\item{\bf Higgs sector}

The Higgs doublet is written as: 
\be \phi=\left(
\begin{array}{c}
i \chi^+ \\
{1 \over \sqrt{2}}(\psi-i \chi_3)
\end{array}
\right)   \label{higgsdoublet}
\end{equation}
where $\chi^\pm={1 \over \sqrt{2}}(\chi^1 \mp \chi^2)$. Among
these fields only $\psi$ has a zero mode, and we assign to such a
mode a vacuum expectation value ${\hat v}$, so that $\psi \to
{\hat v}+H$. $H$ can be identified with the SM Higgs field, while the
vacuum expectation value in 5 dimensions is related to the
corresponding quantity in 4 dimensions through the relation:
$\hat v={v / \sqrt{2 \pi R} }$. 

\item{\bf Mixing between Higgs fields and gauge bosons}

The charged $W_{5(n)}^{\pm }$ and $\chi^\pm_{(n)}$ fields mix,  as well as
the neutral $Z_{5(n)}$ and $\chi_{3(n)}$. The resulting fields
are $G_{(n)}^0$, $G_{(n)}^\pm$ which are Goldstone modes giving
mass to the $W_{(n)}^{\pm \mu}$ and $Z_{(n)}^\mu$,  and $a_{(n)}^0$,
$a_{(n)}^\pm$,  new physical scalars.

\item{\bf  Yukawa terms}

In order to have chiral fermions as in the SM,  the
left and the right-handed components of a given spinor cannot be
simultaneously even under $P_5$. The
Yukawa coupling of  the Higgs  field to the fermions provides the fermion mass
terms, and the diagonalization of such terms leads to
the introduction of the CKM matrix,  with the same steps  as in
 the SM. In this respect,
  the ACD  model belongs to the class of {\it
minimal flavour violation} models, since there are no
new operators beyond those present in the SM and no new phases
beyond the CKM phase. As a  consequence, Buras and
collaborators have shown that the unitarity
triangle is the same as in the SM  \cite{Buras:2002ej}. 
In order to obtain 4-d  mass eigenstates for the higher KK levels, a 
further mixing is
introduced among the left-handed doublet and the right-handed
singlet for each flavour $f$. The mixing angle is such that $\displaystyle \tan (2 \alpha_{f(n)})= {m_f \over n/R}$
($n \ge 1$)  giving the masses $\displaystyle m_{f(n)}=\sqrt{m_f^2+{n^2 \over R^2}}$, so that it
is negligible  for all flavours except  the top.

\end{itemize}

As a result of the construction, the four-dimensional Lagrangian, obtained integrating over the 5th dimension $y$:
\be
{\cal L}_4(x)= \int_0^{2 \pi R} {\cal L}_5(x,y)
\ee
describes
i) zero modes (corresponding to the Standard Model fields),
ii) their massive KK excitations,  iii) KK excitations without zero modes (they do not correspond to any field in SM).  
The related Feynman rules can be found in \cite{Buras:2002ej}.

\section{Decays $ B \to K^{(*)} \ell^+ \ell^-$ } \label{sec:modes}

In the Standard Model the effective $ \Delta B =-1$, $\Delta S = 1$ Hamiltonian
governing  the rare transition $b \to s
\ell^+ \ell^-$ can be written in terms of a set of local
operators:
\begin{equation}
H_W\,=\,4\,{G_F \over \sqrt{2}} V_{tb} V_{ts}^* \sum_{i=1}^{10}
C_i(\mu) O_i(\mu) \label{hamil}
\end{equation}
\noindent where $G_F$ is the Fermi constant and $V_{ij}$ are
elements of the Cabibbo-Kobayashi-Maskawa mixing matrix; we
neglect terms proportional to $V_{ub} V_{us}^*$ since the ratio
$\displaystyle \left |{V_{ub}V_{us}^* \over V_{tb}V_{ts}^*}\right
|$ is of the order $10^{-2}$. The operators $O_i$, written in
terms of quark and gluon fields, read as follows:
\begin{eqnarray}
O_1&=&({\bar s}_{L \alpha} \gamma^\mu b_{L \alpha})
      ({\bar c}_{L \beta} \gamma_\mu c_{L \beta}) \nonumber \\
O_2&=&({\bar s}_{L \alpha} \gamma^\mu b_{L \beta})
      ({\bar c}_{L \beta} \gamma_\mu c_{L \alpha}) \nonumber \\
O_3&=&({\bar s}_{L \alpha} \gamma^\mu b_{L \alpha})
      [({\bar u}_{L \beta} \gamma_\mu u_{L \beta})+...+
      ({\bar b}_{L \beta} \gamma_\mu b_{L \beta})] \nonumber \\
O_4&=&({\bar s}_{L \alpha} \gamma^\mu b_{L \beta})
      [({\bar u}_{L \beta} \gamma_\mu u_{L \alpha})+...+
      ({\bar b}_{L \beta} \gamma_\mu b_{L \alpha})] \nonumber \\
O_5&=&({\bar s}_{L \alpha} \gamma^\mu b_{L \alpha})
      [({\bar u}_{R \beta} \gamma_\mu u_{R \beta})+...+
      ({\bar b}_{R \beta} \gamma_\mu b_{R \beta})] \nonumber \\
O_6&=&({\bar s}_{L \alpha} \gamma^\mu b_{L \beta})
      [({\bar u}_{R \beta} \gamma_\mu u_{R \alpha})+...+
      ({\bar b}_{R \beta} \gamma_\mu b_{R \alpha})] \nonumber \\
O_7&=&{e \over 16 \pi^2} m_b ({\bar s}_{L \alpha} \sigma^{\mu \nu}
     b_{R \alpha}) F_{\mu \nu} \nonumber \\
O_8&=&{g_s \over 16 \pi^2} m_b \Big[{\bar s}_{L \alpha}
\sigma^{\mu \nu}
      \Big({\lambda^a \over 2}\Big)_{\alpha \beta} b_{R \beta}\Big] \;
      G^a_{\mu \nu} \nonumber\\
O_9&=&{e^2 \over 16 \pi^2}  ({\bar s}_{L \alpha} \gamma^\mu
     b_{L \alpha}) \; {\bar \ell} \gamma_\mu \ell \nonumber \\
O_{10}&=&{e^2 \over 16 \pi^2}  ({\bar s}_{L \alpha} \gamma^\mu
     b_{L \alpha}) \; {\bar \ell} \gamma_\mu \gamma_5 \ell
\label{eff}
\end{eqnarray}
\noindent 
where $\alpha$, $\beta$ are colour indices, $\displaystyle
b_{R,L}={1 \pm \gamma_5 \over 2}b$, and $\displaystyle \sigma^{\mu
\nu}={i \over 2}[\gamma^\mu,\gamma^\nu]$; $e$ and $g_s$ are the
electromagnetic and the strong coupling constant, respectively, while
$F_{\mu \nu}$ and $G^a_{\mu \nu}$ in $O_7$ and $O_8$ denote the
electromagnetic and the gluonic field strength tensor. $O_1$ and
$O_2$ are current-current operators, $O_3,...,O_6$ are  QCD
penguin operators, $O_7$ (inducing the radiative $b \to s \gamma$
decay) and $O_8$ are magnetic penguin operators, $O_9$ and
$O_{10}$ are semileptonic electroweak penguin operators.

The Wilson coefficients appearing in (\ref{hamil}) have been
computed at NNLO in the Standard Model  \cite{nnlo}.  At
 NLO  the coefficients have been computed also for  the ACD model including the effects of KK modes
\cite{Buras:2002ej,Buras:2003mk}: we 
 use these results in our study. No operators other than those collected in eq.(\ref{eff}) are found in
 ACD, therefore the whole contribution of the plethora of states only produces a modification
 of the Wilson coefficients that now depend on the  additional ACD parameter, the compactification radius. For large values of $1/R$ the Standard Model phenomenology should be
  recovered, since the new states,  being more and more massive,  decouple from the low-energy theory. 
  Our aim is 
 to establish a  lower bound on $1/R$ from the various $B \to K^{(*)} \ell^+ \ell^-$ observables. 

In the following we do not consider  
 the contribution to $B \to K^* \ell^+ \ell^-$  with the  lepton pair  coming from
 $c{\bar c}$ resonances, which is mainly due to the operators $O_1$ and $O_2$ in (\ref{eff}). It can be 
 experimentally removed   applying 
appropriate kinematical cuts around the resonances. 
  QCD penguins $O_3,...,O_6$ can also be
neglected since their Wilson coefficients are very small compared
to the others. Therefore, we only need the coefficients $C_7$,
$C_9$ and $C_{10}$: as discussed  in
\cite{Buras:2002ej,Buras:2003mk},  the  impact of the KK modes 
consists in the enhancement of $C_{10}$ and the
suppression of $C_7$.

The Wilson  coefficients in ACD are modified
because particles  not present in  SM can contribute
as intermediate states in penguin and box diagrams. As a
consequence, the Wilson coefficients can be expressed in terms of
functions $F(x_t,1/R)$, $x_t=\displaystyle{m_t^2 \over M_W^2}$,
which generalize the corresponding SM functions $F_0(x_t)$
according to: \be
F(x_t,1/R)=F_0(x_t)+\sum_{n=1}^{\infty}F_n(x_t,x_n)
\label{functions} \ee where $x_n=\displaystyle{m_n^2 \over
M_W^2}$ and $m_n=\displaystyle{n \over R}$. The 
relevant functions are the following: $C(x_t,1/R)$  from $Z^0$
penguins; $D(x_t,1/R)$ from $\gamma$ penguins; $E(x_t,1/R)$ from
gluon penguins; $D^\prime (x_t,1/R)$ from $\gamma$ magnetic
penguins; $E^\prime (x_t,1/R)$ from chromomagnetic penguins.
The functions can be found in
\cite{Buras:2002ej,Buras:2003mk}; here we collect the formulae
needed in our analysis.
\begin{itemize}
\item{\bf $C_7$}

In place of $C_7$,  one defines an effective coefficient
$C_7^{(0)eff}$ which is renormalization scheme independent
\cite{Buras:1993xp}: \be C_7^{(0)eff}(\mu_b)=\eta^{16 \over 23}
C_7^{(0)}(\mu_W)+{8 \over 3} \left( \eta^{14 \over 23} -\eta^{16
\over 23} \right)C_8^{(0)}(\mu_W)+C_2^{(0)}(\mu_W) \sum_{i=1}^8
h_i \eta^{\alpha_i} \ee where $\displaystyle \eta={\alpha_s(\mu_W) \over
\alpha_s(\mu_b)}$, and \be C_2^{(0)}(\mu_W)=1 \hskip 0.5cm
C_7^{(0)}(\mu_W)=-{1 \over 2} D^\prime(x_t,1/R), \hskip 0.5cm
C_8^{(0)}(\mu_W)=-{1 \over 2} E^\prime(x_t,1/R); \ee  the
superscript $(0)$ stays for leading log approximation.
Furthermore: \bea a_1={14 \over 23} \hskip 0.4cm a_2={16 \over 23}
\hskip 0.4cm && a_3={6 \over 23} \hskip 0.4cm a_4=-{12 \over 23}
\nonumber \\ a_5= 0.4086 \hskip 0.4cm a_6=-0.4230 \hskip 0.4cm &&
a_7=-0.8994 \hskip 0.4cm
a_8=0.1456 \nonumber \\
h_1=2.2996 \hskip 0.4cm h_2=-1.0880  \hskip 0.4cm &&
 h_3=-{3 \over 7}  \hskip 0.4cm  h_4=-{1 \over 14} \label{numbers}
 \\
 h_5=-0.6494 \hskip 0.4cm h_6=-0.0380 \hskip 0.4cm && h_7= -0.0185 \hskip 0.4cm
 h_8=-0.0057 \,\,\, . \nonumber \eea

The functions $D^\prime$ and $E^\prime $ are given by eq.
(\ref{functions}) with \bea D^\prime_0(x_t)=-{(8 x_t^3+5 x_t^2-7
x_t) \over 12 (1-x_t)^3}+{x_t^2(2-3 x_t) \over 2(1-x_t)^4}\ln x_t  \\
E^\prime_0(x_t)=-{x_t(x_t^2-5 x_t-2) \over 4 (1-x_t)^3} +{3 x_t^2
\over 2 (1-x_t)^4}\ln x_t\eea
\newpage
\bea D^\prime_n(x_t,x_n)&=& {x_t(-37+44 x_t+17 x_t^2+6 x_n^2(10-9
x_t+3 x_t^2) -3 x_n (21-54 x_t+17 x_t^2)) \over 36
(x_t-1)^3 }\nonumber \\
&+&{x_n(2-7 x_n+3 x_n^2) \over 6} \ln {x_n \over 1+ x_n} \nonumber
\\ &-&{(-2+x_n+3 x_t)(x_t+3 x_t^2+x_n^2(3+x_t)-x_n(1+(-10+x_t)x_t))
\over 6 (x_t-1)^4} \ln{x_n+x_t \over 1+x_n} \nonumber \\ \\
E^\prime_n(x_t,x_n)&=&{x_t(-17-8 x_t+x_t^2-3 x_n(21-6 x_t+x_t^2)-6
x_n^2(10-9
x_t+3 x_t^2)) \over 12 (x_t-1)^3} \label{d-eprime} \nonumber \\
&-&{1 \over 2}x_n(1+x_n)(-1+3 x_n)\ln {x_n \over 1+ x_n}\nonumber \\
&+&
{(1+x_n)(x_t+3 x_t^2+x_n^2(3+x_t)-x_n(1+(-10+x_t)x_t)) \over
2(x_t-1)^4} \ln{x_n+x_t \over 1+x_n} \,\,\, . 
 \eea
Following \cite{Buras:2002ej} one gets the expressions for the sum over $n$: 
\bea
\sum_{n=1}^{\infty}D^\prime_n(x_t,x_n)&=& -{x_t(-37 +x_t(44+17
x_t)) \over 72 (x_t-1)^3}\nonumber \\
+ {\pi M_W R \over 2}&& \Bigg[ \int_0^1 dy \, {(2 y^{1/2}+7
y^{3/2}+3 y^{5/2}) \over 6} \coth (\pi M_WR \sqrt{y}) \nonumber \\
&+&{(-2+3 x_t)x_t(1+3 x_t) \over 6 (x_t-1)^4}J(R,-1/2)\nonumber \\
&-& {1 \over 6 (x_t-1)^4} \left[x_t(1+3 x_t)-(-2+3
x_t)(1+(-10+x_t)x_t) \right] J(R, 1/2)\nonumber \\&+& {1 \over 6
(x_t-1)^4} \left[(-2+3
x_t)(3+x_t)-(1+(-10+x_t)x_t) \right]J(R, 3/2)\nonumber \\
&-&{(3+x_t) \over 6 (x_t-1)^4} J(R,5/2)\Bigg] \label{sumn} \\
 \sum_{n=1}^{\infty}E^\prime_n(x_t,x_n)&=&-{x_t(-17+(-8+x_t)x_t)
 \over 24 (x_t-1)^3} \nonumber \\
 +
{\pi M_W R \over 4}&& \Bigg[\int_0^1 dy \, (y^{1/2}+2
y^{3/2}-3 y^{5/2}) \,  \coth (\pi M_WR \sqrt{y}) \nonumber \\
&-&{x_t(1+3 x_t) \over (x_t-1)^4}J(R,-1/2)\nonumber \\
&+& {1 \over  (x_t-1)^4} \left[x_t(1+3 x_t)-(1+(-10+x_t)x_t) \right] J(R, 1/2)\nonumber \\
&-& {1 \over  (x_t-1)^4} \left[(3+x_t)-(1+(-10+x_t)x_t) \right]J(R, 3/2)\nonumber \\
&+&{(3+x_t) \over  (x_t-1)^4} J(R,5/2)\Bigg] \eea where 
 \be J(R,\alpha)=\int_0^1 dy \, y^\alpha \left[ \coth (\pi
M_W R \sqrt{y})-x_t^{1+\alpha} \coth(\pi m_t R \sqrt{y}) \right] \,\,\, .\ee

\item{\bf $C_9$}

In the ACD model and in the NDR scheme one has \be
C_9(\mu)=P_0^{NDR}+{Y(x_t,1/R) \over \sin^2 \theta_W} -4
Z(x_t,1/R)+P_E E(x_t,1/R) \ee where $P_0^{NDR}=2.60 \pm 0.25$
\cite{Misiak:1992bc} and the last term is numerically negligible.
Besides \bea Y(x_t,1/R)&=&Y_0(x_t)+\sum_{n=1}^\infty C_n(x_t,x_n)
\nonumber \\
Z(x_t,1/R)&=& Z_0(x_t)+\sum_{n=1}^\infty C_n(x_t,x_n) \eea with
\bea Y_0(x_t)&=&{x_t \over 8} \left[ {x_t -4 \over x_t -1}+{3 x_t
\over (x_t-1)^2} \ln x_t \right]\,\,\nonumber \\
Z_0(x_t)&=& {18 x_t^4-163 x_t^3+259 x_t^2 -108
x_t \over 144 (x_t-1)^3} \nonumber \\
&+& \left[{32 x_t^4-38 x_t^3-15 x_t^2+18 x_t \over 72
(x_t-1)^4}-{1 \over 9}\right] \ln x_t \,\, \eea
 \be\label{cn}
C_n(x_t,x_n)=\frac{x_t}{8 (x_t-1)^2} \left[x_t^2-8 x_t+7+(3 +3
x_t+7 x_n-x_t x_n)\ln \frac{x_t+x_n}{1+x_n}\right] \ee and \bea &&
\sum_{n=1}^\infty C_n(x_t,x_n)= \nonumber \\&& {x_t(7-x_t) \over
16 (x_t-1)} -{\pi M_W R x_t \over 16 (x_t-1)^2}
\left[3(1+x_t)J(R,-1/2)+(x_t-7)J(R,1/2) \right] \,\,\, . \eea

 \item{\bf
$C_{10}$}

 $C_{10}$ is $\mu$ independent and is given by \be C_{10}=-{Y(x_t,1/R) \over \sin^2
\theta_W}  \,\,\,\, .\ee
\end{itemize}
We  fix the renormalization scale to $\mu=\mu_b \simeq 5 $ GeV.

With these coefficients and the operators in (\ref{eff}) the inclusive $b \to s \ell^+ \ell^-$ transitions
have been studied in \cite{Buras:2002ej,Buras:2003mk}. The exclusive $B \to K^{(*)} \ell^+ \ell^-$
modes, on the other hand, involve  
  the matrix elements of the operators  in
(\ref{eff}) between the $B$ and  $K$ or $K^*$ mesons, for which we  use the
standard parametrization in terms of form factors:

\begin{equation}
<K(p^\prime)|{\bar s} \gamma_\mu b |B(p)>=(p+p^\prime)_\mu
F_1(q^2) +{M_B^2-M_K^2 \over q^2} q_\mu \left
(F_0(q^2)-F_1(q^2)\right ) \label{f0}
\end{equation}
\noindent ( $q=p-p^\prime$, $F_1(0)=F_0(0)$) and
\begin{equation}
<K(p^\prime)|{\bar s}\; i\;  \sigma_{\mu \nu} q^\nu b |B(p)>=
\Big[(p+p^\prime)_\mu q^2 -(M_B^2-M_K^2)q_\mu\Big] \; {F_T(q^2)
\over M_B+M_K} \hskip 3 pt ; \label{ft}
\end{equation}
\begin{eqnarray}
<K^*(p^\prime,\epsilon)|{\bar s} \gamma_\mu (1-\gamma_5) b
|B(p)>&=& \epsilon_{\mu \nu \alpha \beta} \epsilon^{* \nu}
p^\alpha p^{\prime \beta}
{ 2 V(q^2) \over M_B + M_{K^*}}  \nonumber \\
&-& i \left [ \epsilon^*_\mu (M_B + M_{K^*}) A_1(q^2) -
(\epsilon^* \cdot q) (p+p')_\mu  {A_2(q^2) \over (M_B + M_{K^*}) }
\right. \nonumber \\
&-& \left. (\epsilon^* \cdot q) {2 M_{K^*} \over q^2}
\big(A_3(q^2) - A_0(q^2)\big) q_\mu \right ] \label{a1}
\end{eqnarray}
\noindent and
\begin{eqnarray}
<K^*(p^\prime,\epsilon)|{\bar s} \sigma_{\mu \nu} q^\nu
{(1+\gamma_5) \over 2} b |B(p)>&=& i \epsilon_{\mu \nu \alpha
\beta} \epsilon^{* \nu} p^\alpha p^{\prime \beta}
\; 2 \; T_1(q^2)  + \nonumber \\
&+&  \Big[ \epsilon^*_\mu (M_B^2 - M^2_{K^*})  -
(\epsilon^* \cdot q) (p+p')_\mu \Big] \; T_2(q^2) \nonumber \\
&+& (\epsilon^* \cdot q) \left [ q_\mu - {q^2 \over M_B^2 -
M^2_{K^*}} (p + p')_\mu \right ] \; T_3(q^2)  \; . \nonumber \\ \label{t1}
\end{eqnarray}
\noindent $A_3$ can be written as a combination of $A_1$
and $A_2$:
\begin{equation}
A_3(q^2) = {M_B + M_{K^*} \over 2 M_{K^*}}  A_1(q^2) - {M_B -
M_{K^*} \over 2 M_{K^*}}  A_2(q^2)
\end{equation}
with the condition $ A_3(0) = A_0(0)$. The identity $\displaystyle
\sigma_{\mu \nu} \gamma_5 = - {i \over 2}
 \epsilon_{\mu \nu \alpha \beta} \sigma^{\alpha \beta}$
($ \epsilon_{0 1 2 3}=+1$) implies that $T_1(0) = T_2(0)$.

The form factors are non-perturbative quantities. We use for them
 two sets of results:  the first one,  denoted as set A,  
obtained by  three-point QCD sum rules based on the short-distance expansion
\cite{Colangelo:1995jv};  the second one, denoted as set B,   obtained  by  QCD sum rules 
based on the light-cone expansion
\cite{Ball:2004rg}. In both cases, the mass of the $b$ quark is finite.
The $q^2$ dependence is fitted in the region where the methods can be relibly applied,
actually the low $q^2$ region, and then it is extrapolated to the full physical range.
The form factors in set A are fitted with by two different functional dependences:
either polar or linear. $F_1$, $V$ and $T_1$ display a polar behavior,
while   $A_1$, $T_2$, and $A_2$, 
$T_3$ depend linearly on $q^2$, with decreasing (increasing) behaviour. Only  the form factor
$F_T$ is a double pole.
The values of parameters with their estimated errors can be found in \cite{Colangelo:1995jv};
the uncertainties  have been included in the analysis we present below.
In set B of form factors, $A_1$ and
$T_2$ are fitted as simple poles, $V$ and $T_1$ as a sum of two polar
functions and $F_1$, $F_T$, $A_2$, $T_3$ are the sum of a pole and a double
pole.  The values of the parameters and of their estimated uncertainties can be found
in \cite{Ball:2004rg}; also for this set we include in the numerical analysis the errors of the parameters.
The main
differences between the two sets of form factors  essentially concern $A_1$ and
$T_2$.

In the  numerical analyses
  we use the values reported by the PDG
\cite{Eidelman:2004wy} for masses and CKM matrix elements.  We also
use $ m_b=4.8$ GeV, $m_t=172.7$ GeV, coinciding with the central
value recently reported by the Tevatron electroweak working group
\cite{unknown:2005cc},  and $\tau_{B^0}=1.527 \pm 0.008$ ps
\cite{hfag}.

\subsection{$ B \to K \ell^+ \ell^-$}

\begin{figure}[ht]
\begin{center}
\includegraphics[width=0.45\textwidth] {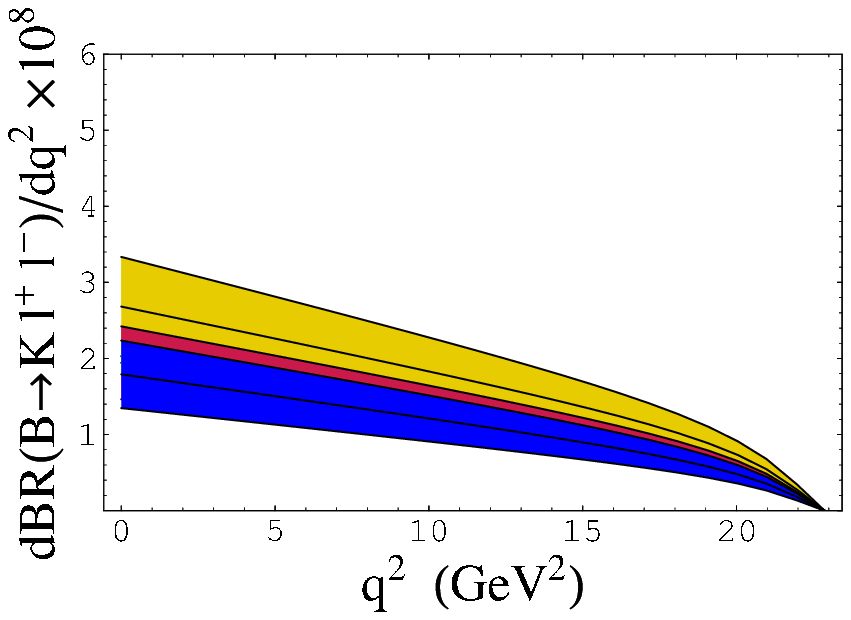} \hspace{0.5cm}
 \includegraphics[width=0.45\textwidth] {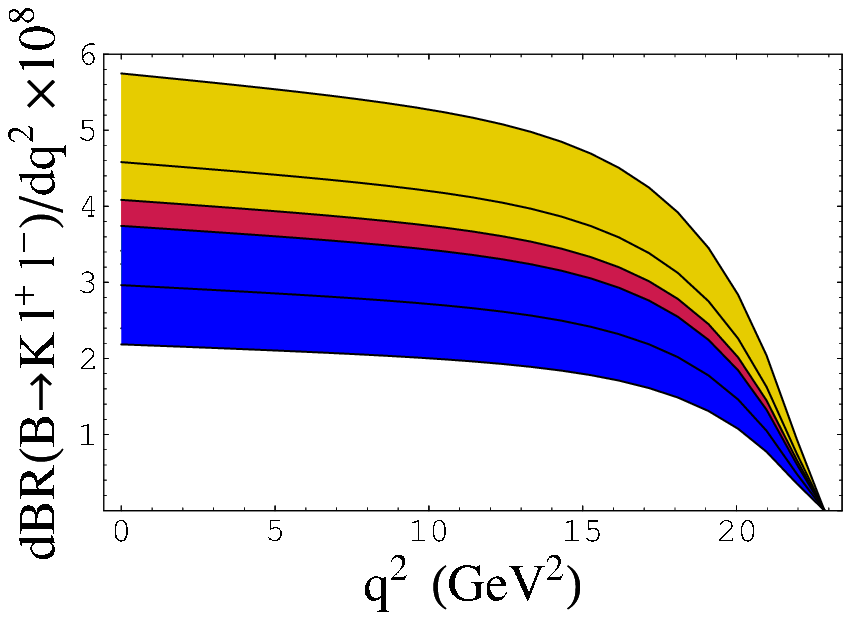}
\end{center}
\caption{\baselineskip=15pt  Differential branching fraction $\displaystyle{dBR(B \to
K \ell^+ \ell^-) / dq^2}$  obtained using  set A   (left) and B  (right) of form factors. The  dark (blue)
region is obtained in  SM;
 the intermediate (red) one for $1/R=500$ GeV, the light (yellow) one for $1/R=200$ GeV.  The contribution of
 $c \bar c$ resonances is not displayed.} \vspace*{1.0cm}
\label{spettrokll}
\end{figure}

\begin{figure}[ht]
\begin{center}
\includegraphics[width=0.45\textwidth] {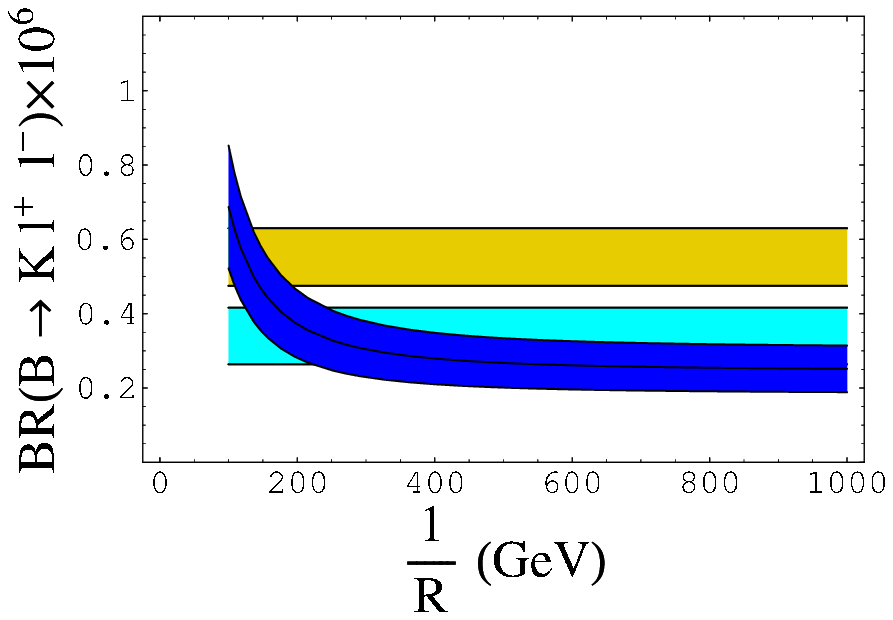} \hspace{0.5cm}
 \includegraphics[width=0.45\textwidth] {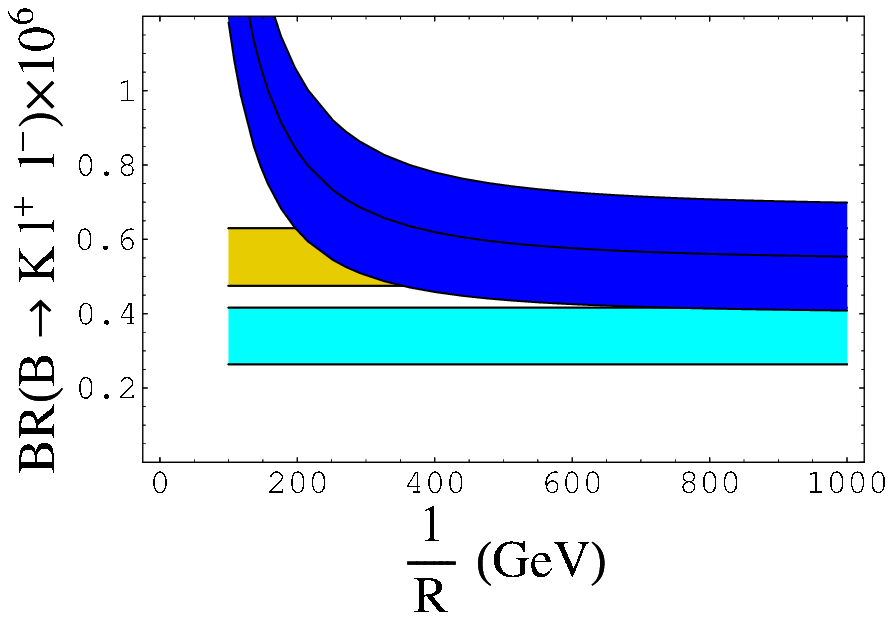}
\end{center}
\caption{\baselineskip=15pt  $BR(B \to K \ell^+ \ell^-)$  versus $\displaystyle{1
\over R}$ using set A (left) and B 
 (right) of form factors.  The two horizontal regions correspond
to the experimental data provided by BaBar (lower band) and Belle
(upper band), see Table  \ref{brbsleptons}.} \vspace*{1.0cm} \label{brkcol}
\end{figure}

For each set of form factors the differential decay rate in the invariant mass
squared of the lepton pair
\begin{eqnarray}
{d \Gamma \over d q^2}(B \to K \ell^+ \ell^-)&=&{M_B^3 G_F^2
\alpha^2 \over
1536 \pi^5} |V^*_{ts} V_{tb}|^2 \times \nonumber \\
&&\left\{ \left|C_7\,2 m_b \left (- {F_T(q^2) \over M_B+M_K}
\right )+ C_9 F_1(q^2)\right |^2+
\left | C_{10} F_1(q^2)\right |^2 \right\} \times \nonumber \\
&& \left[ \left ( 1-{M_K^2 \over M_B^2} \right )^2 + \left ({q^2
\over M_B^2}\right )^2-2 \left( {q^2 \over M_B^2} \right) \left
(1+{M_K^2 \over M_B^2} \right) \right ]^{3/2} \label{spettro_k}
\end{eqnarray}
\noindent ($q^2=M^2_{\ell^+ \ell^-}$)
displays a  dependence on the compactification 
parameter $1/R$, as depicted in Fig. \ref{spettrokll},
where we have considered the values $1/R=200, \,\, 500$ GeV and the case
of the Standard Model (large values of $1/R$).
The maximum effect is
 in the low $q^2$ region, where the spectrum has the maximum.
  However, such an effect is obscured by the hadronic uncertainty,
 which is evaluated  comparing the two set of form factors and taking into account their errors.  
 Theferore, the differential decay rate does not seem the
 most suitable observable for studying the effect of extra-dimension  at the current level of
 hadronic uncertainties. The situation is different for the width.
 In Fig. \ref{brkcol} we
plot, for the two sets of form factors,  the branching fraction as a function of $1/R$ 
 and compare it with the experimental data provided by
BaBar and Belle .
A  constraint cannot be put on $1/R$ if one adopts set A, while set B
 allows to exclude  $1/R \leq 200$ GeV. It is interesting to observe that, within the Standard Model, 
set A prefers the lowest experimental range,  corresponding to the BaBar result, while
set B is in better agreement with Belle data. Improved  measurements
should resolve the present discrepancy between the two  experiments. At the same time,
they should reather easily allow to  increase the lower bound for the compactification parameter.

\subsection{$ B \to K^* \ell^+ \ell^-$}

A great deal of information can be obtained from the mode  $B
\to K^* \ell^+ \ell^-$ by investigating, together with the lepton
invariant mass distribution, also the forward-backward asymmetry
${\cal A}_{fb}$ in the dilepton angular distribution, which may
reveal effects beyond the Standard Model that could not be
seen in the analysis of the decay rate. In particular, 
 in  SM,  due to the opposite sign of the Wilson
coefficients $C_7$ and $C_9$, ${\cal A}_{fb}$ has a zero the
position of which is almost independent of the model for  the form
factors  \cite{Burdman:1998mk}.

Let us define $\theta_\ell$ as the angle between the $\ell^+$
direction and the $B$ direction in the rest frame of the lepton
pair (we consider  massless leptons). The decay
amplitude can be written as sum of non interfering helicity
amplitudes, the double differential decay rate reads as follows:
\begin{eqnarray}
{d^2 \Gamma \over dq^2 dcos\theta_\ell}&=&{G_F^2|V_{tb}V_{ts}^*|^2
\alpha^2 \over 2^{13} \pi^5} {\lambda^{1/2}(M_B^2,M_{K^*}^2,q^2)
\over M_B^3} \times
\left\{ sin^2 \theta_\ell A_{L} + \right.\nonumber \\
&+& \left. q^2\left [(1+cos \theta_\ell)^2(A^L_+ + A^R_-) + (1-cos
\theta_\ell)^2(A^L_- + A^R_+)\right]\right\} \label{dg2}
\end{eqnarray} \noindent where $A_{L}$ corresponds to a
longitudinally polarized $K^*$, while $A^{L(R)}_{+(-)}$ represent
the contribution from left (right) leptons and from $K^*$ with
transverse polarization: $\displaystyle\epsilon_\pm=\Big(0,{1
\over \sqrt{2}},\pm {i \over \sqrt{2}},0 \Big)$:
\begin{equation}
A_{L}={1 \over M_{K^*}^2} \left\{\left|B_1(M_B^2-M_{K^*}^2-q^2) +
B_2 \lambda\right|^2+ \left|D_1(M_B^2-M_{K^*}^2-q^2) + D_2
\lambda\right|^2\right\} \label{along} \end{equation} \noindent
and
\begin{equation}
A^L_\pm=|\lambda^{1/2}(A-C) \mp (B_1-D_1)|^2 \label{left}
\end{equation}
\begin{equation}
A^R_\pm=|\lambda^{1/2}(A+C) \mp (B_1+D_1)|^2 \; ,\label{right}
\end{equation} \noindent where
$\lambda=\lambda(M_B^2,M_{K^*}^2,q^2)$. The terms
$A$,$C$,$B_1$,$D_1$ contain Wilson  coefficients and
form factors:
\begin{equation}
A={C_7 \over q^2}\,4\, m_b\, T_1(q^2)+C_9\,{V(q^2) \over M_B +
M_{K^*}}
 \label{a}
\end{equation}
\begin{equation}
C= C_{10}\, {V(q^2) \over M_B+M_{K^*}} \label{c} \end{equation}
\begin{equation}
B_1={C_7 \over q^2}\, 4\, m_b\, T_2(q^2)(M_B^2-M_{K^*}^2)+ C_9\,
A_1(q^2) ( M_B + M_{K^*}) \label{b1}
\end{equation}
\begin{equation}
B_2=- \left [{C_7 \over q^2}\, 4\, m_b\, \left ( T_2(q^2)+q^2
{T_3(q^2) \over (M_B^2-M_{K^*}^2)} \right )+ C_9 {A_2(q^2)\over
M_B + M_{K^*}} \right ] \label{b2}
\end{equation}
\begin{equation}
D_1= C_{10}\,A_1(q^2)\,(M_B+M_{K^*}) \label{d1} \end{equation}
\begin{equation}
D_2=- C_{10} \,{A_2(q^2) \over M_B+M_{K^*}} \; . \label{d2}
\end{equation} The forward-backward asymmetry,  defined as
\begin{equation}
A^{FB} (q^2)=\displaystyle {\displaystyle \int_0^1{d^2 \Gamma
\over dq^2 d cos\theta_\ell}dcos\theta_\ell -\int^0_{-1}{d^2
\Gamma \over dq^2 d cos\theta_\ell}dcos\theta_\ell
\over\displaystyle \int_0^1{d^2 \Gamma \over dq^2 d
cos\theta_\ell}dcos\theta_\ell +\int^0_{-1}{d^2 \Gamma \over dq^2
d cos\theta_\ell}dcos\theta_\ell} \; ,
 \label{asim_def}
\end{equation}
\noindent reads
\begin{equation}
A^{FB}(q^2)={3\over 4}{ 2 q^2 (A^L_+ +A^R_- -A^L_- -A^R_+) \over
A_{L} + 2 q^2 (A^L_- + A^R_+ + A^L_+ + A^R_-)} \; .\label{asim}
\end{equation}

\begin{figure}[ht]
\begin{center}
\includegraphics[width=0.45\textwidth] {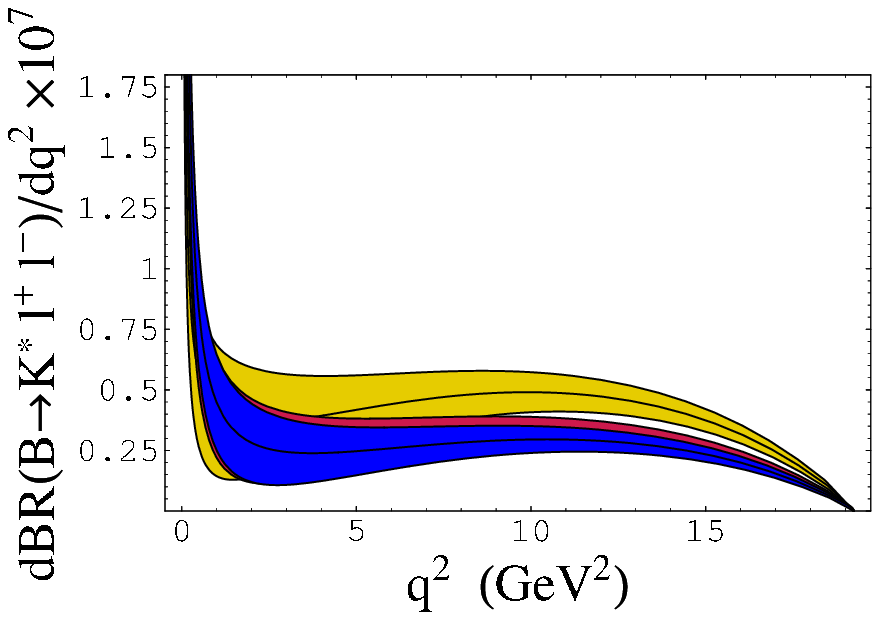} \hspace{0.5cm}
 \includegraphics[width=0.45\textwidth] {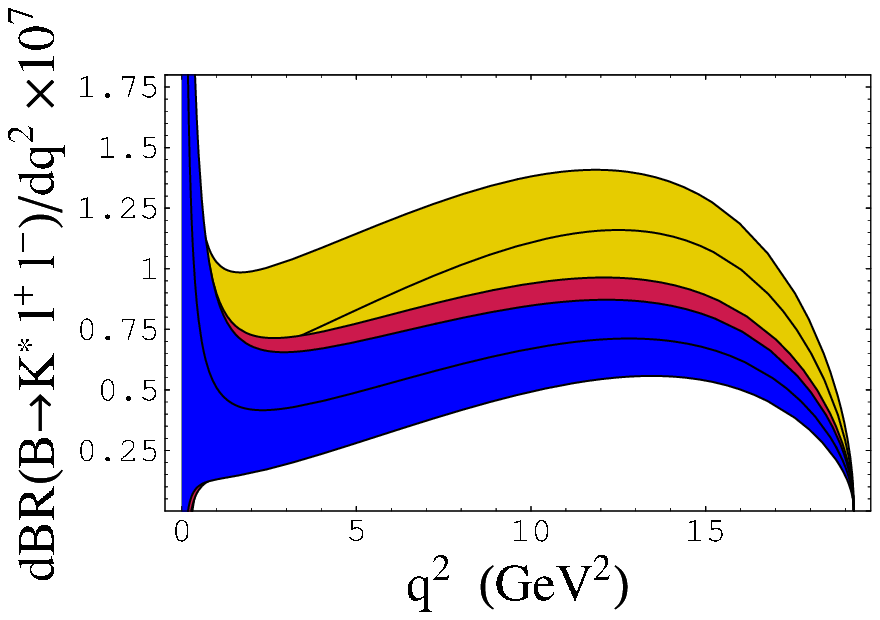}
\end{center}
\caption{\baselineskip=15pt Differential branching fraction $\displaystyle{dBR(B \to
K^* \ell^+ \ell^-) / dq^2}$  using set A (left) and B  (right) of form factors. The dark (blue)
region is obtained in the SM,
 the intermediate (red) one for $1/R=500$ GeV, the light (yellow) one for $1/R=200$ GeV. } \vspace*{1.0cm}
\label{spettrokstarll}
\end{figure}

\begin{figure}[ht]
\begin{center}
\includegraphics[width=0.45\textwidth] {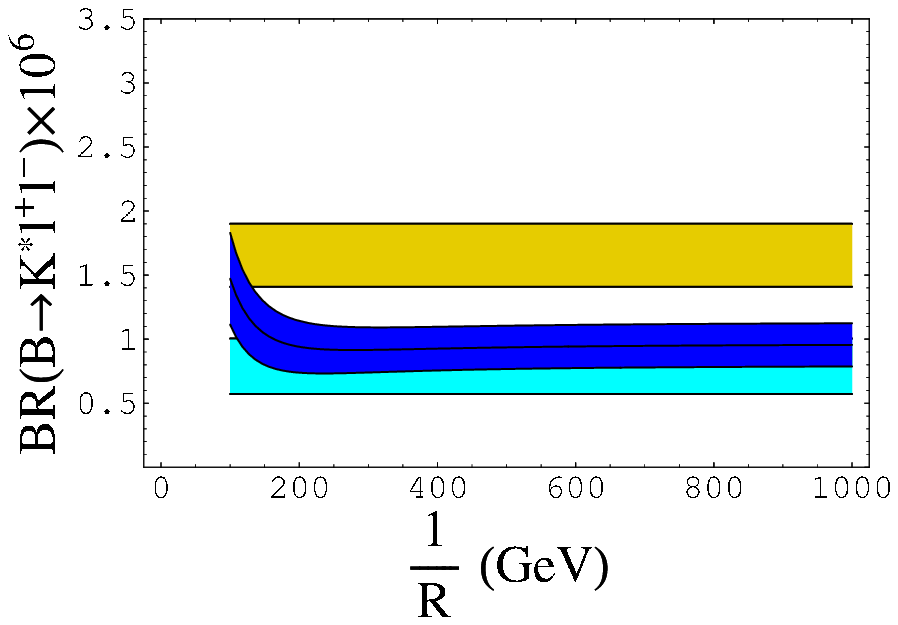} \hspace{0.5cm}
\includegraphics[width=0.45\textwidth] {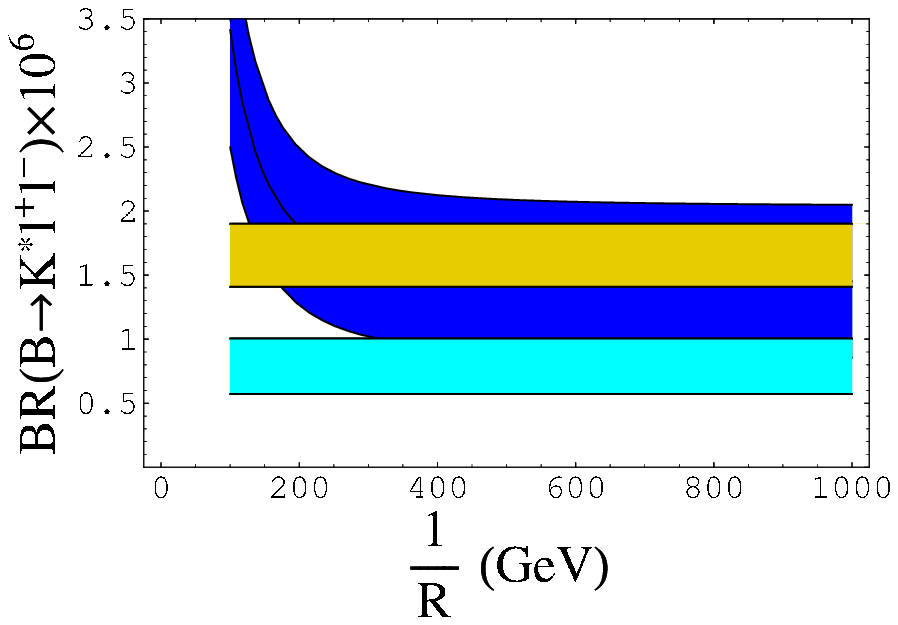}
\end{center}
\caption{\baselineskip=15pt $BR(B \to K^* \ell^+ \ell^-)$  versus $\displaystyle{1
\over R}$ using set A (left) and B (right). The two horizontal regions correspond
to  BaBar (lower band) and Belle (upper band)  results, see Table   \ref{brbsleptons}. } \vspace*{1.0cm} \label{brkstar}
\end{figure}

We can now discuss the predictions of  the ACD model for
the branching ratio as well as for the lepton forward-backward asymmetry.
The differential
branching ratio is shown in Fig. \ref{spettrokstarll}. As in
the case of $B \to K \ell^+ \ell^-$. 
the spectrum is enhanced for lower values of $1/R$,  however, due to  the hadronic uncertainty,
it is not possible to clearly disentangle the  extra dimension  effect. As for the
 total rate,  depicted in  in  Fig. \ref{brkstar} for the two sets of form
factors, set A does not allow to establish a lower bound on $1/R$, while for set B
one gets again $1/R>200$ GeV. The present discrepancy between BaBar and Belle measurements
does not permit  stronger statements, more precise data from both the experiments are expected.

\begin{figure}[ht]
\begin{center}
\includegraphics[width=0.45\textwidth] {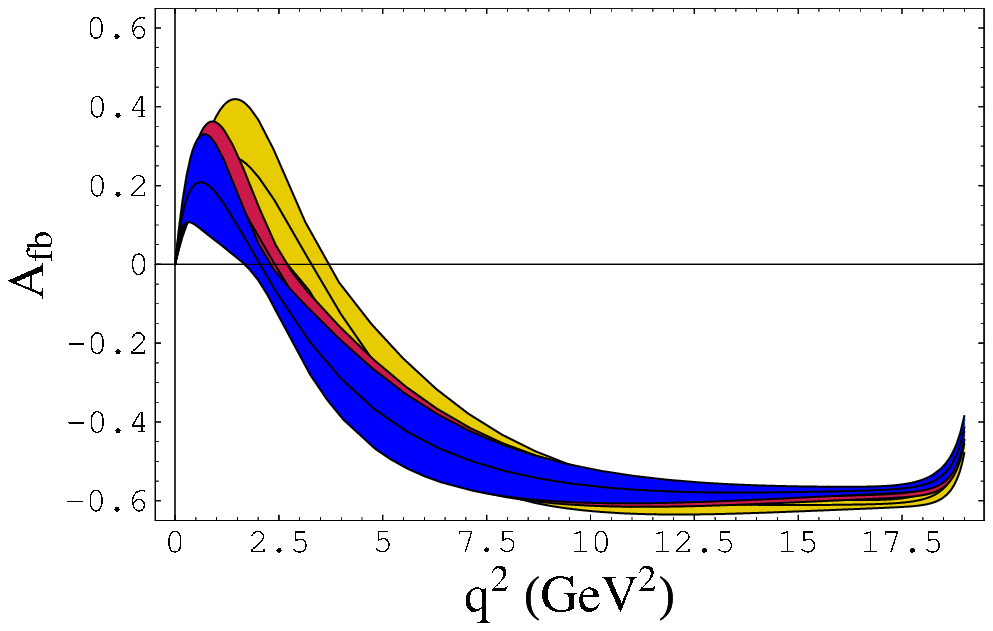} \hspace{0.5cm}
\includegraphics[width=0.45\textwidth] {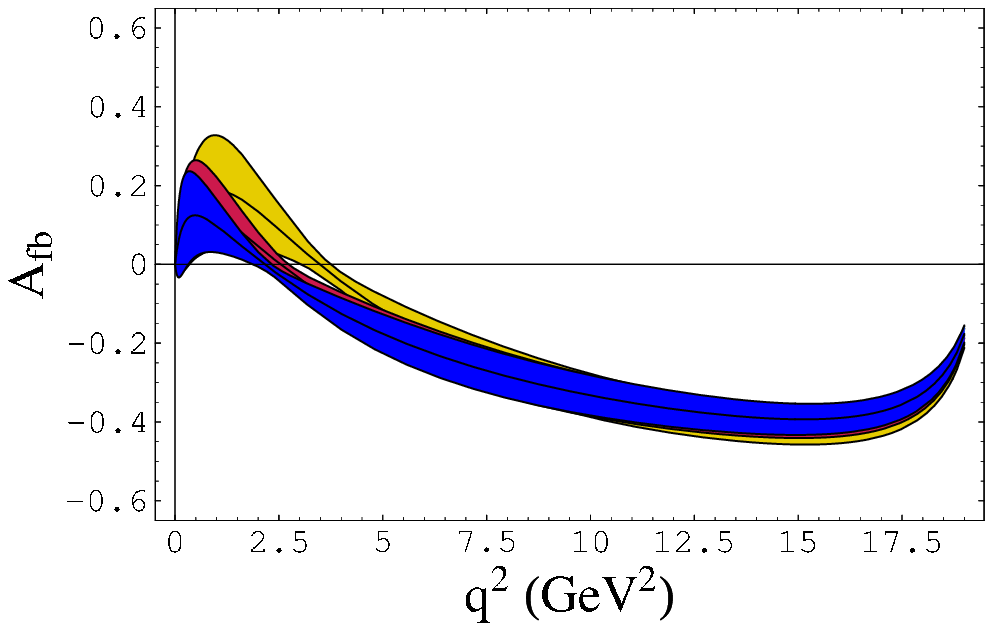}
\end{center}
\caption{\baselineskip=15pt Forward-backward lepton
asymmetry in $B \to K^* \ell^+ \ell^-$  versus $\displaystyle{1 \over
R}$ using set A  (left) and B (right). The light (yellow) bands correspond to the SM
results,  the intermediate (red) band to $1/R=250$ GeV, the dark (blue) one to
 $1/R=200$ GeV. } \vspace*{1.0cm} \label{fig-afb}
\end{figure}

More information comes from  the forward-backward
asymmetry. We show in Fig. \ref{fig-afb} the 
 predictions for the SM,  $1/R=250$ GeV  and
$ 1/R=200$ GeV. The zero of ${\cal A}_{fb}$ is  sensitive
 to the compactification parameter, so that  its experimental determination  would  constrain $1/R$.

%
We can elaborate on this point, since it is  easy to see, using  (\ref{along})-(\ref{d2}) in
(\ref{asim}),  that the position of the zero of the
asymmetry, $s_0$, is determined by the equation: \be Re(C_9)+{2m_b \over
s}C_7 \left[ (M_B+M_{K^*}){T_1(s) \over V(s)}
+(M_B-M_{K^*}){T_2(s) \over A_1(s)} \right]=0
\,\,\,.\label{zeroasim} \ee
It is noticeable that, 
in the large energy limit of the final state light vector meson, a model independent prediction for the position of the zero of the
asymmetry can be obtained. As a matter of fact, in this 
 limit the form factors of  $B \to P,V$ obey spin
symmetry relations \cite{Charles:1998dr}, broken by hard gluon
corrections to the weak vertex and hard spectator interactions. In
the heavy-quark limit one can  write \cite{Beneke:2003pa} (see
also \cite{Hill:2002vw,Lange:2003pk,Bauer:2002aj}): \be \langle P|
\bar \psi \, \Gamma_i \, b |B\rangle =  C_i(E, \mu_I) \,
\xi_P(\mu_I,E) +
 T_i(E,u,\omega,\mu_{\rm II}) \otimes \phi_+^B(\omega,\mu_{\rm II}) \otimes
         \phi_P(u,\mu_{\rm II})+
\nn   \ldots, \label{factorization}
 \ee where $\Gamma_i$ is a generic Dirac
structure and the dots stand for sub-leading terms in
$\Lambda/m_b$. In the case of a vector meson $V$ two functions
$\xi_{\perp,\parallel}$ (depending on the Dirac structure $\Gamma_i$) appear in place of $\xi_P$. The matrix
elements in (\ref{factorization}) get therefore two contributions.
The first one contains the short-distance functions
 $C_i$,  arising from integrating
out hard modes: $\mu_I<m_b$, and a  ``soft'' form factor $\xi_P$
which   does not depend on the Dirac structure of the decay
current. The second term in (\ref{factorization}) factorizes into
a hard-scattering kernel $T_i$ and the light-cone distribution
amplitudes $\phi_B$ and $\phi_P$.  A still controversial
question is to what extent the first contribution is numerically
suppressed by Sudakov effects \cite{li}, although it has been put
forward that the
second term in (\ref{factorization}) should be subleading with respect to the first one 
\cite{Lange:2003pk,Descotes-Genon:2001hm,DeFazio:2005dx} . Neglecting  ${\cal O}(\alpha_s)$ effects
the approximate symmetry relations mentioned above between the
vector and tensor form factors for $B \to V$ transitions 
read  \cite{Charles:1998dr,Beneke:2000wa}:

\be {T_1(E) \over V(E)}={1 \over 2}{M_B \over M_B +M_{K^*}} \hskip
1cm {T_2(E) \over A_1(E)}={(M_B +M_{K^*}) \over 2 M_B  }
\label{symrel} \,.\ee
The use of eq. (\ref{symrel}) in (\ref{zeroasim})  produces a
form-factor independent result for the position of the zero of the
asymmetry: using our numerical input parameters, one
would obtain in the Standard Model  $s_0^{LEET}\simeq 3.61$ GeV$^2$. On the other hand,
taking into account corrections to the relations in
(\ref{symrel}), the
position of the zero of ${\cal A}_{fb}$ moves to
$s_0\simeq 4.2 \pm 0.6$ GeV$^2$  \cite{Beneke:2001at}.

\begin{figure}[htb]
\begin{center}
\includegraphics[width=0.45\textwidth] {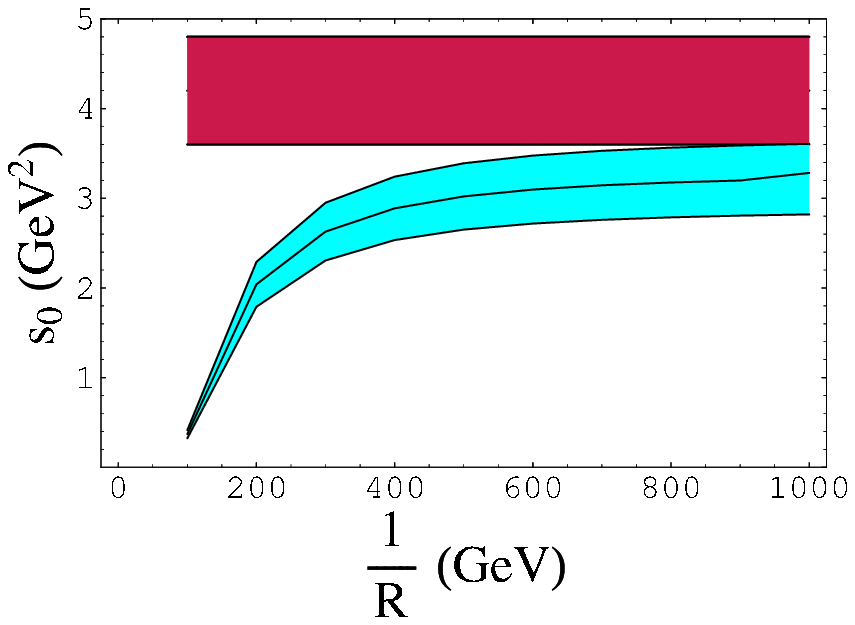} \hspace{0.5cm}
 \includegraphics[width=0.45\textwidth] {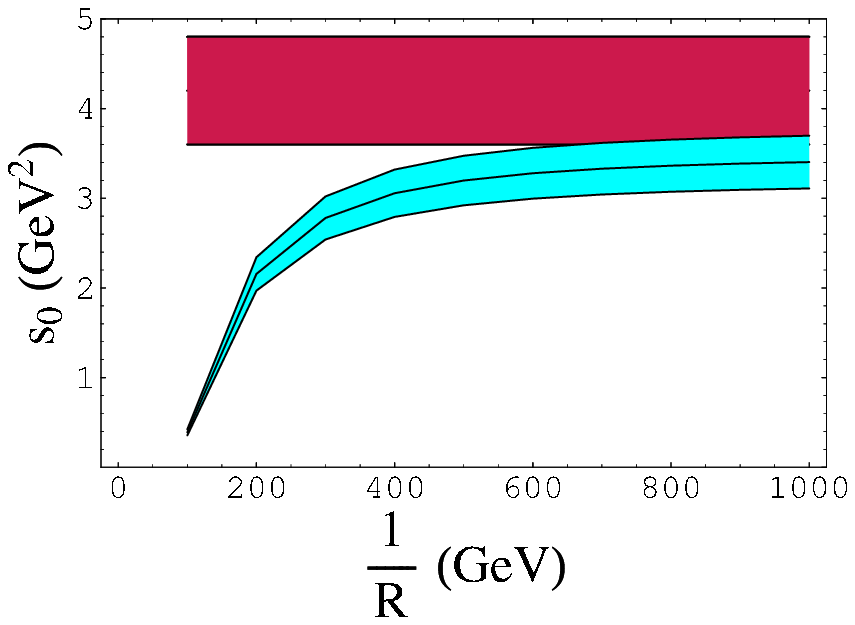}
\end{center}
\caption{\baselineskip=15pt Zero of the forward-backward lepton asymmetry versus
$\displaystyle{1 \over R}$  obtained using the set A (left) anf B (right) of
form factors. In the  plots  the
horizontal region
 represents the  the value of $s_0$ derived in \cite{Beneke:2001at}. } \vspace*{1.0cm} \label{zerocol}
\end{figure}

The dependence of $s_0$ on the compactification parameter is depicted 
in Fig. \ref{zerocol} for the  sets A and B of form factors. Two considerations are in order.
The first one is that the value of $s_0$ in the Standard Model is only marginally consistent
with the result obtained in \cite{Beneke:2001at}, suggesting that further corrections
could shift $s_0$ to smaller values. The second one concerns the sensible dependence
of $s_0$ on the compactification parameter: in particular, the zero is pushed to  low values by
decreasing $1/R$. Such a sensitivity in the exclusive channel is analogous to the one observed
in the inclusive $B \to X_s \ell^+ \ell^-$ mode, and indicates that $s_0$ is particularly suited to constrain $1/R$. 

At present, the analysis of the forward-backward lepton asymmetry performed by Belle Collaboration
indicates that the relative sign of the Wilson coeffcients $C_9$ and $C_7$ is negative, confirming
that ${\cal A}_{fb}$ should have a zero \cite{Abe:2005km}.
The 
accurate measurement of $s_0$ in the exclusive $B \to K^* \ell^+ \ell^-$ channel is therefore
within the reach of current experiments.

\section{The decays $B \to K^{(*)} \nu {\bar \nu}$}
As mentioned in the Introduction, 
among the various flavour changing neutral current-induced
$b-$quark decays the transitions induced by  $ b \to s \nu {\bar \nu} $ play
a peculiar role, both from a theoretical and an experimental point
of view.
Within the Standard Model these processes  are governed by
the effective Hamiltonian \be {\cal H}_{eff} = {G_F \over \sqrt 2}
{\alpha \over 2 \pi \sin^2(\theta_W)} \; V_{ts} V^*_{tb} \; \eta_X X(x_t)
\; {\bar b} \gamma^\mu ( 1- \gamma_5) s \; {\bar \nu} \gamma_\mu (
1- \gamma_5) \nu
 \equiv  c_L^{SM} \; {\cal O}_L
\label{hamilnunubar} \ee obtained from $Z^0$ penguin and box
diagrams where the dominant contribution corresponds to a top
quark intermediate state. In (\ref{hamilnunubar})  $\theta_W$ the
Weinberg angle. ${\cal O}_L$ represents
the left-left four-fermion operator ${\cal O}_L \equiv {\bar b}
\gamma^\mu ( 1- \gamma_5) s \; {\bar \nu} \gamma_\mu ( 1-
\gamma_5) \nu $. The  function $X$ has been computed in
\cite{inami}  and \cite{buchalla,buchalla1};
we put to unity the QCD factor $\eta_X$ \cite{buchalla,buchalla1,Buchalla:1998ba}.
Possible
New Physics (NP) effects  can  modify the SM value of the
coefficient $c_L$, or introduce one new right-right operator: \be
{\cal H}_{eff} \equiv c_L \; {\cal O}_L + c_R \; {\cal O}_R
\label{NP}\ee (${\cal O}_R \equiv {\bar b} \gamma^\mu ( 1+
\gamma_5) s \; {\bar \nu} \gamma_\mu ( 1 + \gamma_5) \nu $), with
$c_R$ only receiving contribution from phenomena beyond SM
\cite{Buchalla:2000sk}.

Another reason of interest for $ b \to s \nu {\bar \nu} $ 
is the absence of long-distance contributions
 related to four-quark operators in the
effective hamiltonian. In this respect, the transition to
neutrinos represents a clean process even in comparison with the
$b \to s \gamma$ decay, where long-distance contributions are
expected to be present, although small \cite{Colangelo:1989gi}.

Within the Standard Model, form factors are needed to predict branching ratios and decay
distributions for the exclusive modes $B \to K^{(*)} \nu {\bar
\nu}$ (see, e.g. \cite{Colangelo:1996ay,Buchalla:2000sk}).
 The inclusion of effects stemming from one universal extra
dimension is straightforward and requires the generalization of
the function $X(x_t)$ \cite{Buras:2002ej}:

\be X(x_t,1/R)=X_0(x_t)+\sum_{n=1}^\infty C_n(x_t,x_n)  \ee where:
\be X_0(x_t)={x_t \over 8} \left[ {x_t+2 \over x_t -1}+{3 x_t -6
\over (x_t -1)^2} \ln x_t \right] \ee and $ C_n (x_t,x_n)$
 defined in eq. (\ref{cn}).

\subsection{$B \to K \nu {\bar \nu}$}

It is interesting to consider the missing energy distribution in the decay $B
\to K \nu {\bar \nu}$. We define $E_{miss}$ the energy of the
neutrino pair in the $B$ rest frame and consider the dimensionless
variable $x=E_{miss}/M_B$, which varies in the range \be {1 -r
\over 2} \le x \le 1-\sqrt{r} \label{xvar} \ee \noindent with
$r=M_K^2/M_B^2$. The differential decay rate is then given by \be
{d \Gamma(B \to K \nu \bar \nu) \over dx} = 3\;{ |c_L + c_R|^2
\,|F_1(q^2)|^2 \over 48 \pi^3 M_B} \sqrt{\lambda^3(q^2, M_B^2,
M_K^2)}\;, \label{scal} \label{e:specK} \ee where $q^2=M_B^2 (2 x-
1)+M_K^2$ and the sum over the three neutrino species is
understood.

\begin{figure}[ht]
\begin{center}
\includegraphics[width=0.45\textwidth] {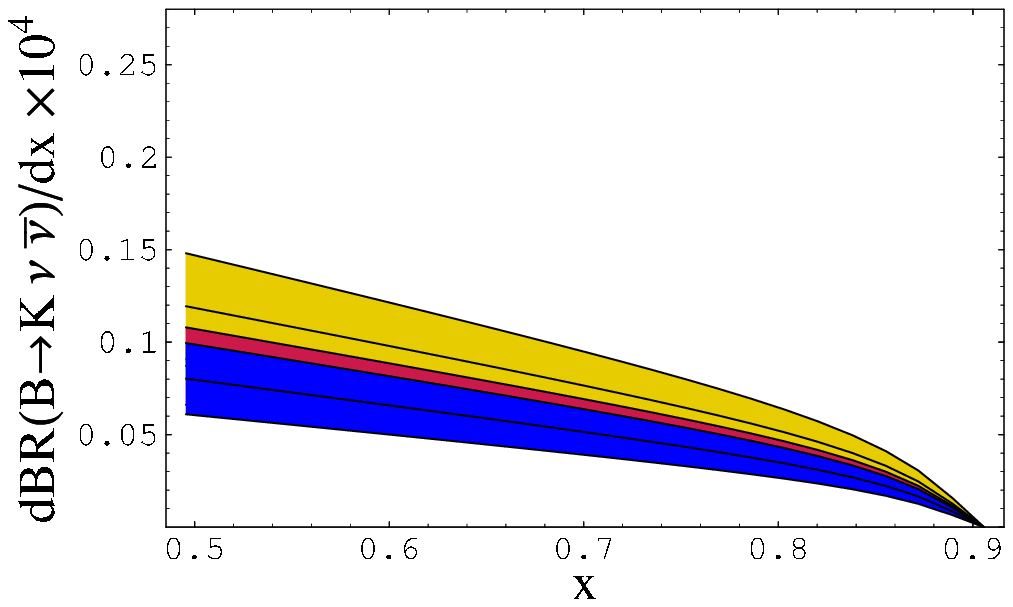} \hspace{0.5cm}
 \includegraphics[width=0.45\textwidth] {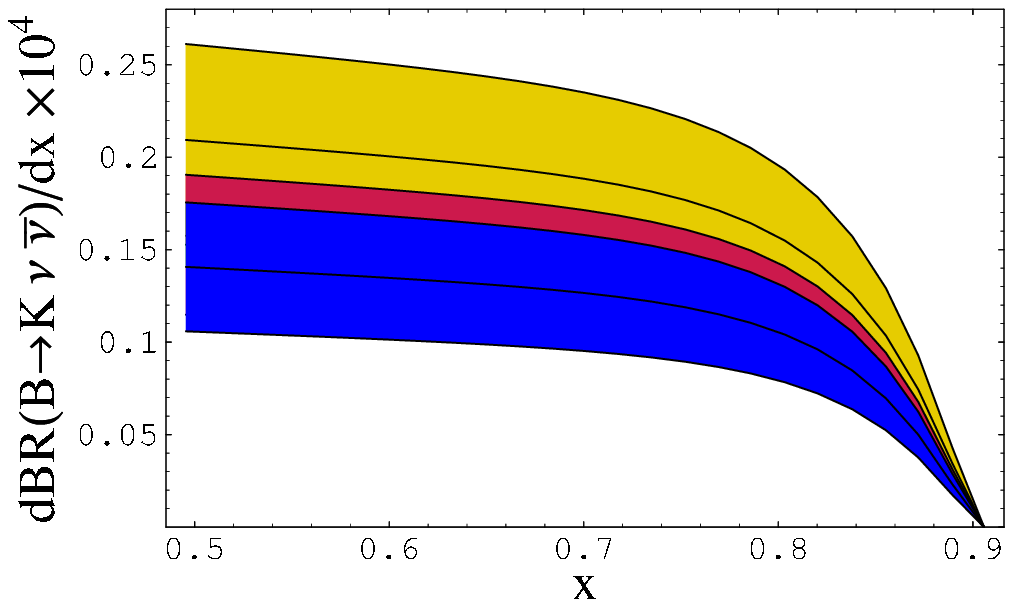}
\end{center}
\caption{\baselineskip=15pt Missing energy distribution in the decay $B \to K \nu
{\bar \nu}$ for set A  (left) and B (right) of form factors.
The sum over the three neutrino species is
understood. 
 The  dark (blue) region is obtained within  SM,
 the intermediate (red) one for $1/R=500$ GeV and the light (yellow) one for $1/R=200$ GeV.} \vspace*{1.0cm}
\label{spectrumbknunu}
\end{figure}
\begin{figure}[ht]
\begin{center}
\includegraphics[width=0.45\textwidth] {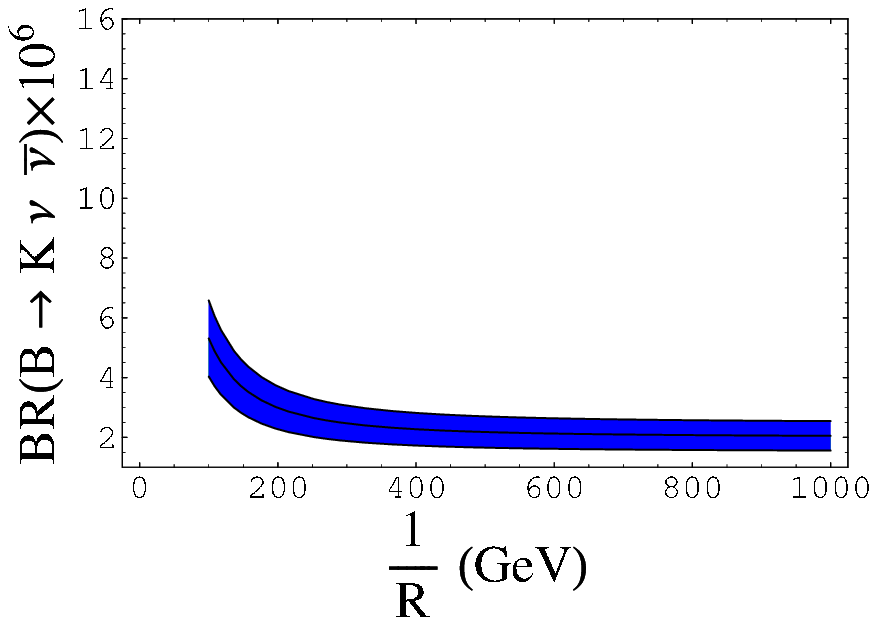} \hspace{0.5cm}
 \includegraphics[width=0.45\textwidth] {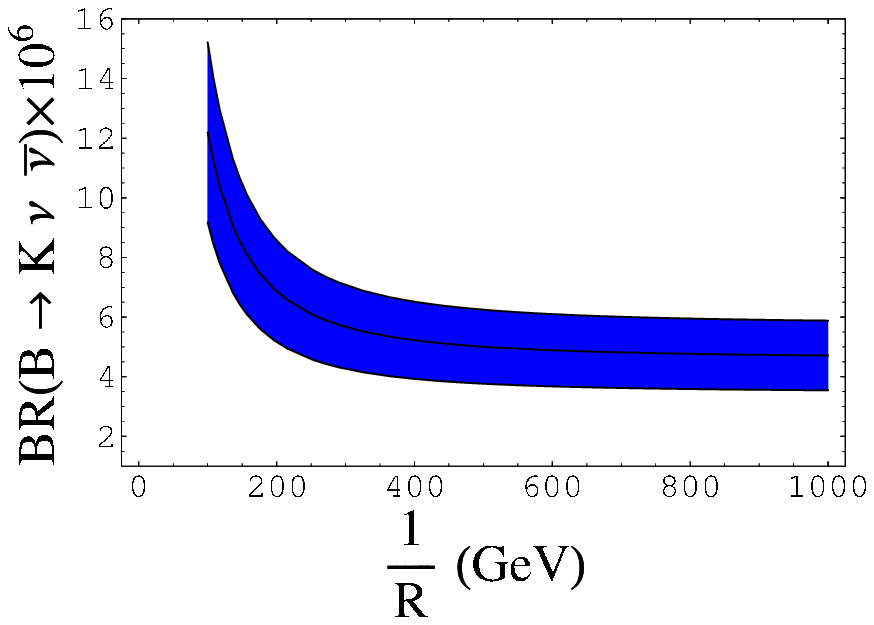}
\end{center}
\caption{\baselineskip=15pt ${BR}(B \to K \nu {\bar \nu})$ versus 1/R for   set A  (left) and  B (right)
 of form factors. } \vspace*{1.0cm}
\label{brbknunu}
\end{figure}
In Fig. \ref{spectrumbknunu} we plot the SM missing energy distribution,  
together with the distributions obtained in  ACD  for different values of $1/R$. 
This distribution is  sensitive to $1/R$, and the largest effect is in the low-x region, however
the hadronic uncertainty is too large for envisaging the possibility of constraining the
compactification parameter. As for the branching fraction,  its dependence on $1/R$ is
shown in Fig. \ref{brbknunu}.   Only an  experimental upper bound exist in
this channel, which is presently too large for any consideration:  however, considering 
Fig. \ref{brbknunu} one sees that the
$1/R$ dependence is too mild for  distinguishing values above $1/R \ge 200$ GeV.

\subsection{$B \to K^* \nu {\bar \nu}$}

\begin{figure}[ht]
\begin{center}
\includegraphics[width=0.45\textwidth] {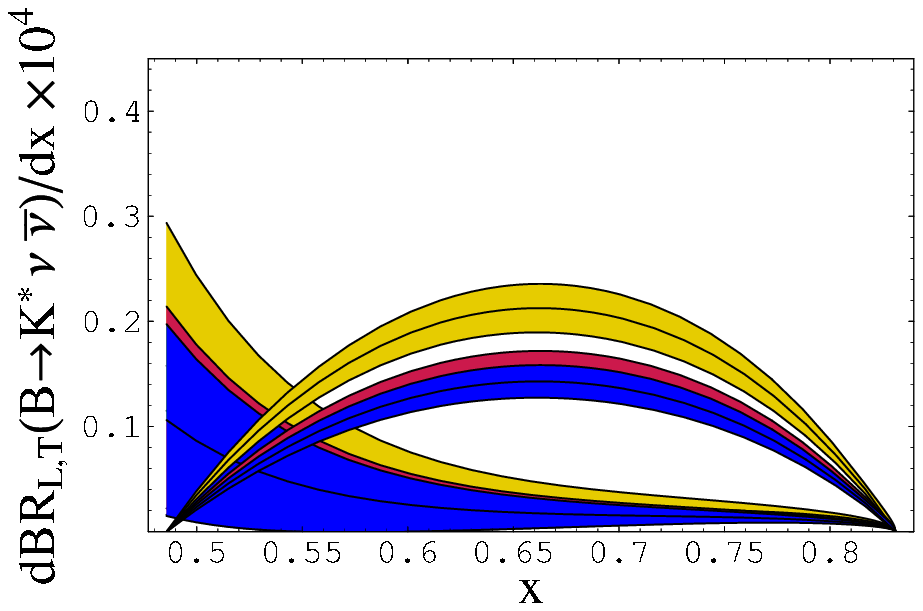} \hspace{0.5cm}
 \includegraphics[width=0.45\textwidth] {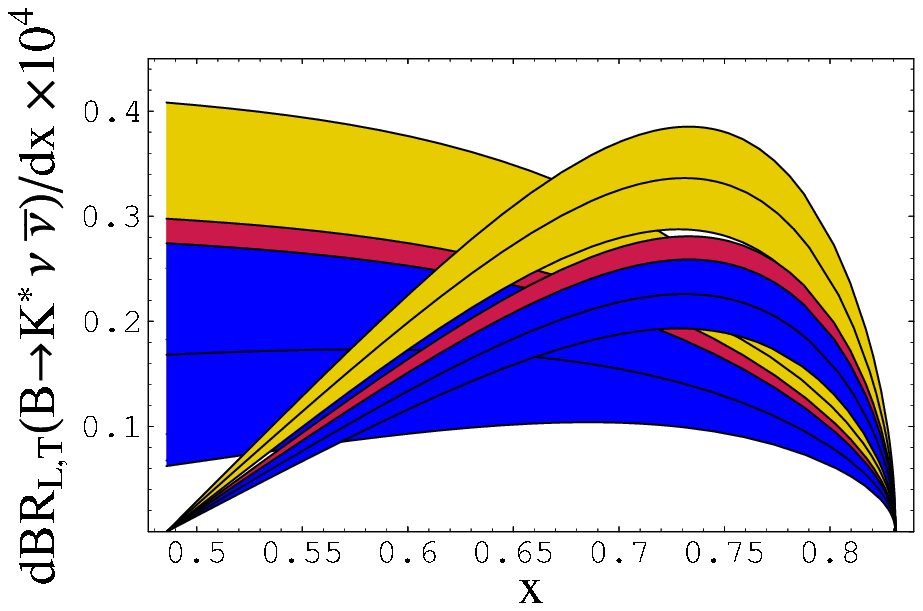}
\end{center}
\caption{\baselineskip=15pt Missing energy distribution in  $B \to K^* \nu
{\bar \nu}$ for a longitudinally polarized $K^*$ (lower curves)
and a transversally polarized $K^*$ (upper curves) for set A (left) and B (right)
of form factors. The sum over
the three neutrino species is understood.  The dark (blue) region corresponds to SM,
 the intermediate (red) one  to $1/R=500$ GeV and  the light (yellow) one  to $1/R=200$ GeV.} \vspace*{1.0cm}
\label{spectrumLTnunu}
\end{figure}

\begin{figure}[ht]
\begin{center}
\includegraphics[width=0.45\textwidth] {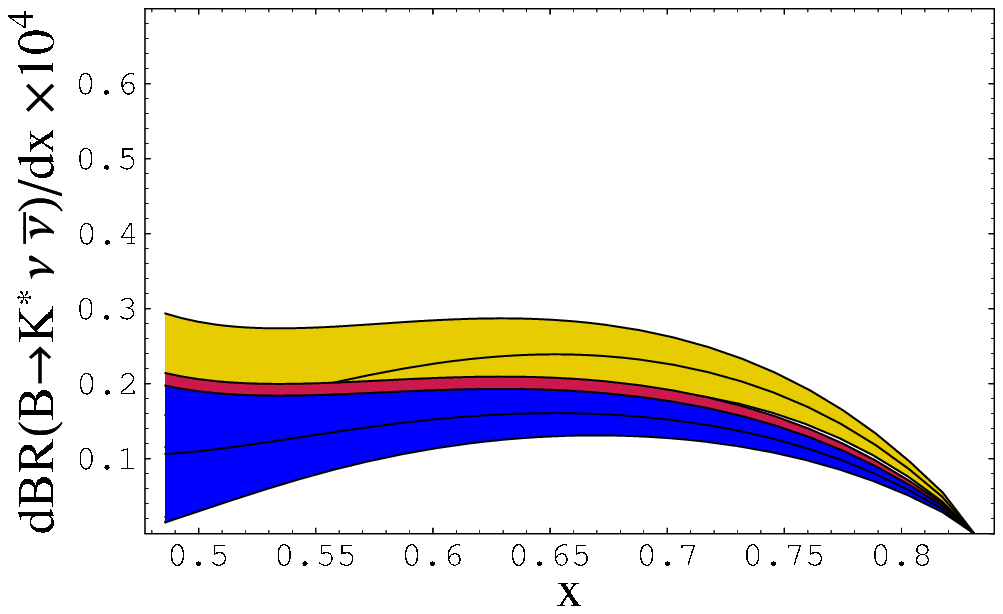} \hspace{0.5cm}
 \includegraphics[width=0.45\textwidth] {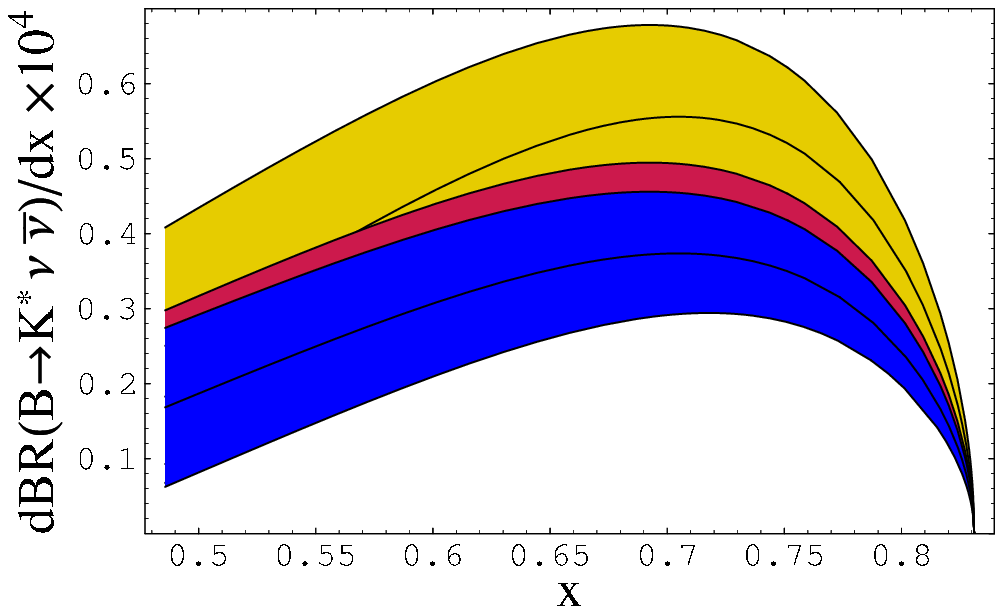}
\end{center}
\caption{\baselineskip=15pt  Missing energy distribution for unpolarized $K^*$,
with the same notations as in Fig. \ref{spectrumLTnunu}.} \vspace*{1.0cm}
\label{spectrumbkstarnunu}
\end{figure}

\begin{figure}[ht]
\begin{center}
\includegraphics[width=0.45\textwidth] {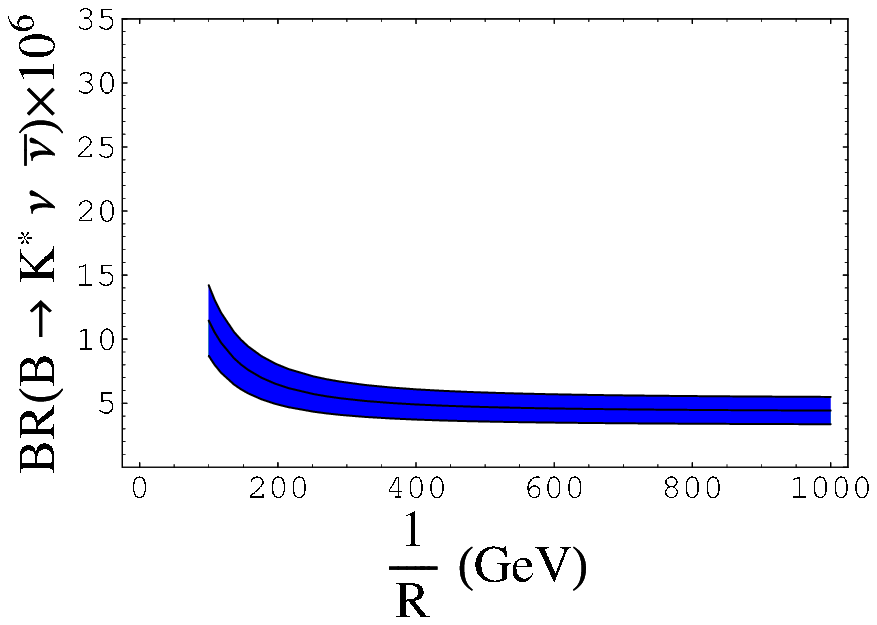} \hspace{0.5cm}
 \includegraphics[width=0.45\textwidth] {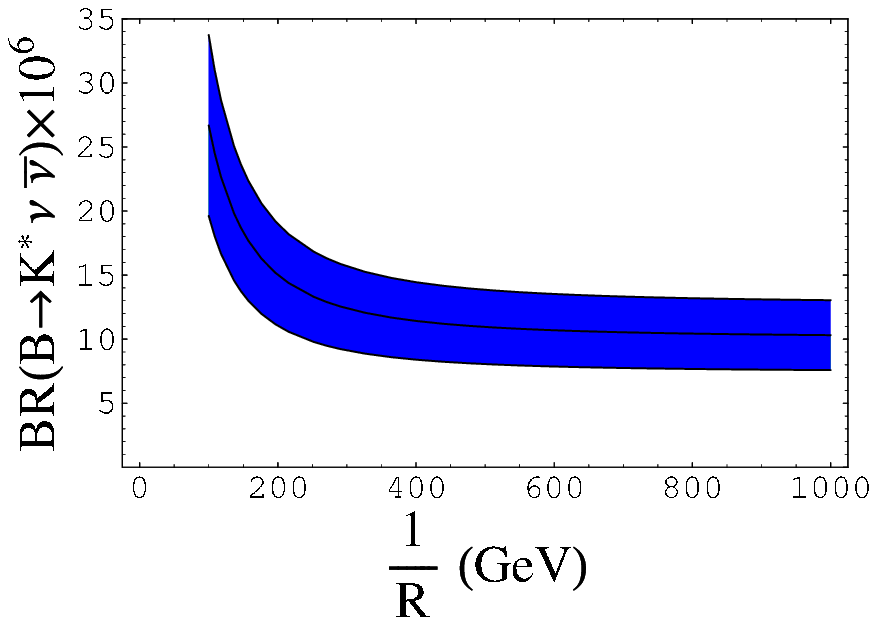}
\end{center}
\caption{\baselineskip=15pt ${BR}(B \to K^* \nu {\bar \nu})$ versus $1/R$
using    set A (left) and B (right) of form factors. } \vspace*{1.0cm}
\label{brbkstarnunu}
\end{figure}

For this mode one can separately consider the missing
energy distributions for  longitudinally and
transversely polarized $K^*$: \be {d \Gamma_L \over dx} = 3\,{
|c_L |^2 \over 24 \pi^3} { |\vec p^{~\prime}| \over M_{K^*}^2}
\left [ (M_B + M_{K^*})(M_B E'-M_{K^*}^2) A_1(q^2) - {2 M_B^2
\over M_B + M_{K^*}} |\vec p^{~\prime}|^2 A_2(q^2) \right ]^2\,,
\label{long} \ee and \be {d \Gamma_{\pm} \over dx} = 3 { |\vec
p^{~\prime}| q^2 \over 24 \pi^3} |c_L |^2 \left |  { 2 M_B |\vec
p^{~\prime}| \over M_B + M_{K^*}} V(q^2) \mp   (M_B + M_{K^*})
A_1(q^2) \right |^2\, \label{tran} \ee where $\vec p^{~\prime}$
and $E'$ are the $K^*$ three-momentum and energy in the $B$ meson
rest frame and the sum over the three neutrino species is understood.
The  missing energy distributions for polarized $K^*$ are shown in Fig.
\ref{spectrumLTnunu}, while the unpolarized  distribution 
and the  branching fraction
are plotted in Fig. \ref{spectrumbkstarnunu}
 and \ref{brbkstarnunu}, respectively. 

Both the polarized $K^*$ and total missing energy distributions depend on $1/R$, but
the hadronic uncertainty obscures such an effect as in $B \to K \nu \bar \nu$. The dependence of the
branching fraction is not strong for  $1/R\ge 200$ GeV.

\section{The decay $ B \to K^* \gamma$ }

As a final case we consider the radiative channel   $ B \to K^* \gamma$, which is the first one observed
among such rare decay modes. 
The transition $b \to s \gamma$ is described  by the operator
$O_7$  in the effective hamiltonian (\ref{eff}), and the $ B \to K^* \gamma$ 
 decay rate is given by: \be \Gamma(B \to K^* \gamma)={\alpha
G_F^2 \over 8 \pi^4}|V_{tb}V_{ts}^*|^2m_b^2 |C_7^{(0)(eff)}|^2
[T_1(0)]^2 M_B^3 \left( 1-{M_{K^*}^2 \over M_B^2} \right)^3 \,\,\, .
\label{ratebkstargamma} \ee

\begin{figure}[ht]
\begin{center}
\includegraphics[width=0.45\textwidth] {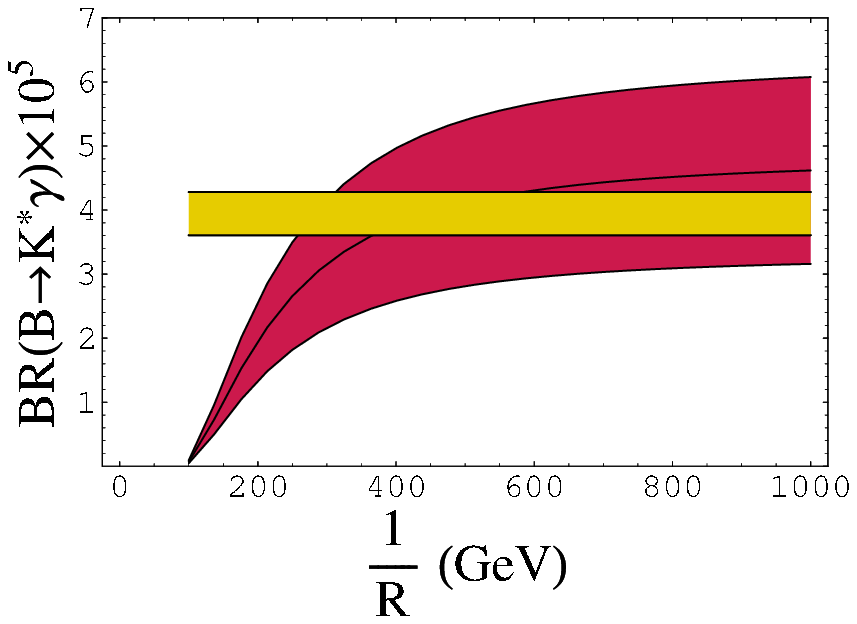} \hspace{0.5cm}
 \includegraphics[width=0.45\textwidth] {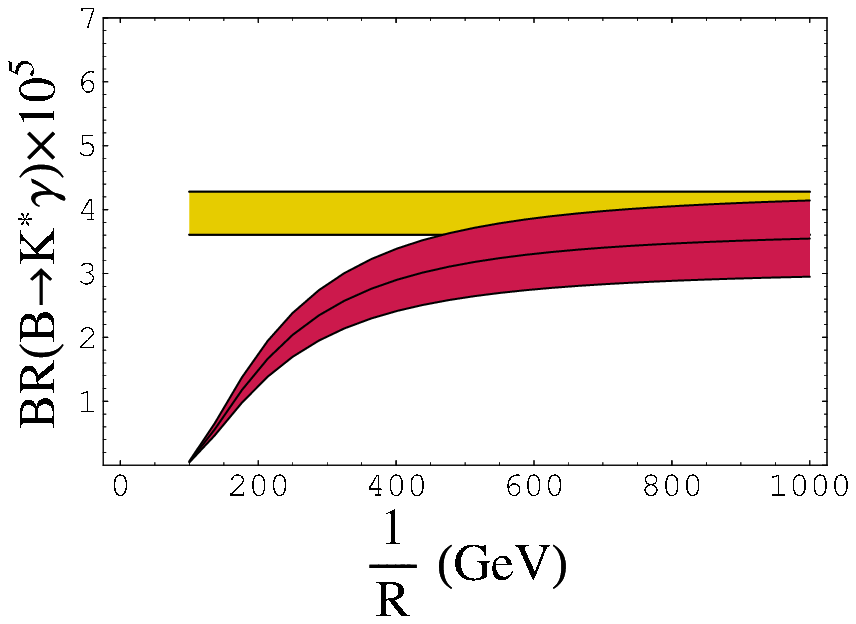}
\end{center}
\caption{\baselineskip=15pt $BR(B \to K^* \gamma)$  versus $\displaystyle{1 \over R}$
using  set A  (left) and   B
 (right) of form factors .  The  horizontal
band corresponds to the experimental result. } \vspace*{1.0cm} \label{brkstargamma}
\end{figure}

One can appreciate the consequences of the existence  of a single
universal extra dimension considering Fig. \ref{brkstargamma}, where the
 branching fraction is plotted as a function of  $1/R$: the sensitivity to the compactification
 parameter is evident, and it allows to put a lower bound of
  $1/R \ge 250$ GeV  adopting set A, and a stronger bound of    $1/R \ge
400$ GeV for set B, which is the most stringent bound that can be currently put on this
parameter from the set of $B$ decay modes we have considered.

\section{Conclusions and Perspectives} \label{sec:concl}
We have analysed the rare $B \to K^{(*)} \ell^+ \ell^-$, $B \to
K^{(*)}\nu \bar \nu$, $B \to K^* \gamma$ decays in the ACD model
with a single universal extra dimension, studying how the
predictions for   branching fractions, decay
distributions and the lepton forward-backward asymmetry  in $B \to K^* \ell^+
\ell^-$ are modified by the introduction of the fifth dimension.
The possibility to constrain the only free parameter of the model,
  the inverse of the compactification radius $1/R$, are
  slightly model dependent, in the sense that the constraints are
different using different sets of form factors. Nevertheless,
 various distributions, together with the lepton
forward-backward asymmetry in $B \to K^* \ell^+ \ell^-$ are very promising in order to constrain $1/R$.
We found that the most stringent lower bound comes from $B \to K^* \gamma$.
Improvements in the experimental data, expected in the near
future, will allow to establish more stringent constraints for the compactification radius.

\vspace*{1cm} \noindent
{\bf Acnowledgments} We thank A.J. Buras for discussions. Two of us (PC and FD) thank CPhT,  \'Ecole
Polytechnique,  for hospitality during the completion of this work.
We acknowledge partial support from the EC Contract No.
HPRN-CT-2002-00311 (EURIDICE).

\end{document}